\documentclass[aps,prb,twocolumn,showpacs,preprintnumbers]{revtex4}
\usepackage{bm,hyperref}
\usepackage{amsmath}
\usepackage{graphicx}
\usepackage{soul}
\usepackage{amsfonts}
\usepackage{amssymb}
\usepackage{dcolumn}
\usepackage{amsmath}

\begin{document}
\title{Fundamental Limits of Optical Force and Torque}

\author{A. Rahimzadegan{$^{*1}$}, R. Alaee{$^{*1,2}$}, I. Fernandez-Corbaton{$^{1,3}$}, and C. Rockstuhl{$^{1,3}$}}
\address{
{$^{1}$}Institute of Theoretical Solid State Physics, Karlsruhe Institute of Technology, Karlsruhe, Germany\\
{$^{2}$}Max Planck Institute for the Science of Light, Erlangen, Germany\\
{$^{3}$}Institute of Nanotechnology, Karlsruhe Institute of Technology, Karlsruhe, Germany\\
$^*$Equally contributed to the work. Corresponding authors: aso.rahimzadegan@kit.edu, rasoul.alaee@mpl.mpg.de
}

\begin{abstract}
Optical force and torque provide unprecedented control on the spatial motion of small particles. A valid scientific question, that has many practical implications, concerns the existence of fundamental upper bounds for 
the achievable force and torque exerted by a plane wave illumination with a given intensity.
Here, while studying isotropic particles, we show that different light-matter interaction channels contribute to the exerted force and torque; and analytically derive upper bounds for
each of the contributions. Specific examples for particles that achieve those upper bounds are provided. We study how and to which extent different contributions can add up to result in the maximum  optical force and torque. Our insights are important for
applications ranging from molecular sorting, particle manipulation, nanorobotics up to ambitious projects such as laser-propelled spaceships.
\end{abstract}
\maketitle
Optical scattering, extinction, and absorption cross sections characterize the strength of light-matter-interaction. They quantify the fraction of power a particle scatters, extincts, or absorbs~\cite{Jackson1999,Bohren2008}.
To describe the interaction of light with a particle, the incident and scattered fields can be expanded into vector spherical wave functions (VSWFs). VSWFs are the eigenfunctions of the vectorial wave equation in spherical coordinates. 
For an isotropic, i.e. a rotationally symmetric particle, the amplitudes of the VSWFs expanding the incident and scattered fields are linked by the Mie coefficients ~\cite{Jackson1999,Xu1995c}. Each coefficient describes a channel
for the light-matter-interaction and is uniquely specified by the total angular momentum (AM) number $j$, and the parity of the fields involved in the scattering process in the respective Mie channel (MC). Depending on $j$ and the parity, 
these MCs are referred to as either electric ($a_j$) or magnetic ($b_j$).

\begin{figure}
\begin{centering}
\includegraphics[scale=0.35]{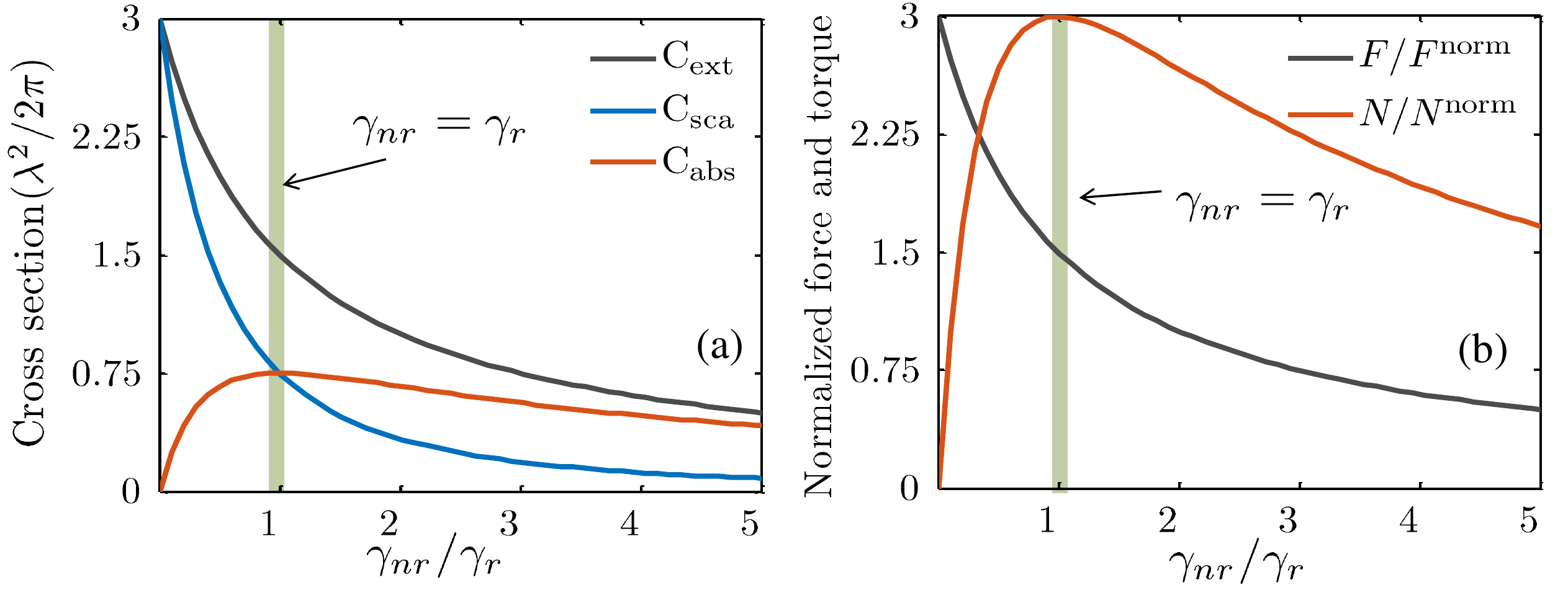}
\par\end{centering}
\protect\caption{(a)  Maximal scattering, extinction, and absorption cross sections of; and (b) optical force and torque exerted on an isotropic dipolar particle as a function of the loss
factor, i.e. $\gamma_\mathrm{nr}/\gamma_\mathrm{r}$. The particle is illuminated by an arbitrarily polarized plane wave (for the cross sections and the force) or a circularly polarized plane wave (only for the torque) at the resonance of a spectrally isolated dipole MC. \label{fig: ED_LossFactor} }
\end{figure}

If an isotropic particle is illuminated by a plane wave in a frequency interval where only a single MC is significant, the maximum scattering cross section $C_\mathrm{sca}$ (at resonance) is $\left(2j+1\right)\lambda^{2}/2\pi$ \cite{ruan2010superscattering,Ruan2011a}. 
The maximum $C_\mathrm{sca}$ is attained when the particle operates in the \textit{over-coupling}
($\gamma_\mathrm{r}\gg\gamma_\mathrm{nr}$) regime, i.e. the radiative (scattering)
loss ($\gamma_\mathrm{r}$) is much larger than the non-radiative (Ohmic)
loss ($\gamma_\mathrm{nr}$). For a particle with a single electric dipole MC (i.e., electric dipolar particle),
the maximum scattering and consequently extinction cross section $C_\mathrm{ext}$ (at resonance) corresponds to $3\lambda^{2}/2\pi$~{[}Fig.~\ref{fig: ED_LossFactor}
(a){]}. Similarly, the maximum absorption cross section $C_\mathrm{abs}$ (at resonance) is $\left(2j+1\right)\lambda^{2}/8\pi$
\cite{miroshnichenko2016ultimate}. It occurs if the particle operates in the \textit{critical coupling} ($\gamma_\mathrm{nr}=\gamma_\mathrm{r}$)
regime, i.e. the non-radiative ($\gamma_\mathrm{nr}$) and radiative ($\gamma_{r}$)
loss are equal. For an electric dipolar particle, the maximum absorption is $3\lambda^{2}/8\pi$~{[}Fig.~\ref{fig: ED_LossFactor}
(a) when $\gamma_\mathrm{nr}=\gamma_\mathrm{r}${]}.

Although optical cross sections are important in studying light-matter interaction
at the nanoscale~\cite{maslovski2016overcoming,cox2002experiment,yang2008superscatterer}, the optical force and torque are further key quantities~\cite{chen2011optical,nieto2015optical,jiang2015poynting,Yangcheng2013giant}, for which upper bounds have not yet been well studied. This is surprising considering the important applications and implications of the optical
force and torque in many areas. Examples are the opto-mechanical manipulation~\cite{dienerowitz2008optical,Grier2003,yannopapas2008optical,rahimzadegan2016optical}, molecular or particle optical sorting~\cite{vcivzmar2006optical},
or nanorobotics~\cite{mavroidis2013nanorobotics,nieminen2005optically}. Optical force and torque are also important in studying the angular momentum of light~\cite{stewart2005angular,franke2008advances,barnett2010rotation,Fernandez-Corbaton2012}. Moreover, developing ambitious projects like laser-propelled
spaceships, would benefit from an understanding of these limits ~\cite{long2011deep}.

In this Letter, based on the multipole expansion in scattering theory~\cite{Bohren2008}, we identify and analyze different terms that contribute to the exerted optical force and torque on isotropic particles; 
and derive upper bounds for each contribution. Next, considering these contributions, the maximum of the total optical force and torque is calculated. Contrary to the
optical cross sections, the force and torque, in a general direction, contain terms that are the result of interference among different MCs.

We start by analyzing particles that are characterized by a single dipole MC. Afterwards, we consider homogeneous dielectric spheres supporting multiple MCs and distinguish different terms contributing 
to the force and torque. Finally, considering more general isotropic particles, the maximum optical force and torque as a function of the maximum non-negligible multipolar order is calculated. Examples for 
particles that maximize each of the contributions as well as the force/torque are given. The detailed derivations of the relations, supplementary figures, and more information on the theoretical 
background are given in a \textit{supplemental
material (SM)}. Here, we concentrate on the presentation of the results and the discussion of the physical implications. The force and torque values are all time averaged.

\textit{Fundamental limits on optical force (electric dipole)}:
An arbitrarily polarized, time harmonic plane wave, propagating in the +z direction illuminates an isotropic electric dipolar particle. 
The exerted force
is \cite{nieto2010optical,chen2011optical}:
\begin{eqnarray}
\mathbf{F}_{\mathrm{p}} = \frac{1}{2}\Re( \nabla\mathbf{E}^{\mathrm{*}}\cdot\mathbf{p}) =\frac{kI_{0}}{c}\,\Im\left[\alpha\left(\omega\right)\right]\,\mathbf{e}_{z}=F_{\mathrm{p}}\mathbf{e}_{z},\label{eq:dipole-Force}
\end{eqnarray} 
\noindent where $\mathbf{p}=\epsilon_{0}\alpha\mathbf{E}$
is the induced Cartesian electric dipole moment, $\alpha$ is the
electric polarizability of the particle, $I_{0}=\epsilon_{0}c|E_{0}|^{2}/2$
is intensity of the illumination, $k$ is the wavenumber, $\epsilon_{0}$
is the free space permittivity, and $c$ is the speed of light. Alternatively, based on the definition of the extinction cross section of an electric
dipolar particle, $C_{\mathrm{ext,p}}=k\mathrm{Im}\left(\alpha\right)$
\cite{tretyakov2003analytical}, $F_\mathrm{p}$ can be rewritten
as $\left(I_{0}/c\right)C_{\mathrm{ext,p}}$.
The dispersion of $\alpha$ near
a resonance can be expressed by a Lorentzian line-shape as~\cite{sipe1974macroscopic}:
\begin{eqnarray}
\alpha\left(\omega\right) & = & \frac{\alpha_{0}}{\omega_{\mathrm{res}}^{2}-\omega^{2}-\mathrm{i}\omega\left(\gamma_\mathrm{nr}+\gamma_\mathrm{r}\right)},\label{eq:alphaEE-Lorenz}
\end{eqnarray}
where $\omega_{\mathrm{res}}$ is the resonance frequency,
$\alpha_{0}$ is the resonance strength, and $\gamma_\mathrm{r}=\alpha_{0}k^{2}/6\pi c$
is the radiative loss of the particle, respectively. The maximum force
in Eq.~\ref{eq:dipole-Force} occurs when the particle is non-absorptive and at resonance (i.e.$\left\{ \mathrm{Im}\left[\alpha\left(\omega\right)\right]\right\} _{\mathrm{max}}=6\pi/k^{3}$), where the extinction cross section is maximized (=$3 \lambda ^2 /2\pi$),
and reads as:
\begin{eqnarray}
\left(F_{\mathrm{p}}\right)_{\mathrm{max}}=3F^{\mathrm{norm}}, &  & F^{\mathrm{norm}}=\frac{I_{0}}{c}\frac{\lambda^{2}}{2\pi}.\label{eq:Fpmax}
\end{eqnarray}

$\left(F_{\mathrm{p}}\right)_{\mathrm{max}}$ is the fundamental limit for the force an arbitrarily polarized plane wave can exert on
an isotropic electric dipolar particle. Due to the
symmetry of Maxwell's equations, the same bound can be attained for an isotropic magnetic dipolar particle. $F^{\mathrm{norm}}$
is used as the normalization factor further on. It is important to note that $\left(F_{\mathrm{p}}\right)_{\mathrm{max}}$  depends on the resonance wavelength ($\propto \lambda^{2}$) of the MC. 
Therefore, the longer the resonance wavelength, the larger the maximum force.

\textit{Fundamental limits on optical torque (electric dipole)}:
If a circularly polarized plane wave $\mathbf{E}=E_{\mathrm{0}}e^{\mathrm{i}kz}\left(\mathbf{e}_{x}+\sigma\mathrm{i}\mathbf{e}_{y}\right)/\sqrt{2}$, with handedness $\sigma=\pm1$, impinges on the particle,
the optical torque reads as \cite{nieto2015optical}:
\begin{eqnarray}
\mathbf{N}_{\mathrm{p}} & = & \frac{1}{2}[\Re(\mathbf{p}\times\mathbf{E}^{\mathrm{*}})-\frac{k^{3}}{6\pi\epsilon_{0}}\Im(\mathbf{p}^{*}\times\mathbf{p})]\nonumber \\
 & = & \frac{\sigma I_{0}}{\omega}\{k\Im\left[\alpha\left(\omega\right)\right]-\frac{k^{4}}{6\pi}\left|\alpha\left(\omega\right)\right|{}^{2}\} \mathbf{e}_{z}=N_{\mathrm{p}}\mathbf{e}_{z}.\label{eq:Np}
\end{eqnarray}

Based on the definition of the absorption cross section of an electric
dipolar particle, $C_{\mathrm{abs,p}}=k\mathrm{Im}\left(\alpha\right)-k^{4}\left|\alpha\right|{}^{2}/6\pi$
\cite{tretyakov2003analytical}, $N_p$ can be rewritten
as $\sigma\left(I_{0}/\omega\right)C_{\mathrm{abs,p}}$.
Therefore, the torque is maximized at the maximum of the
absorption cross section ($=3\lambda^{2}/8\pi$) and is equal to:
\begin{eqnarray}
\left(N_\mathrm{p}\right)_{\mathrm{max}}=3\sigma N^{\mathrm{norm}}, &  & N^{\mathrm{norm}}=\frac{I_{0}}{\omega}\frac{\lambda^{2}}{8\pi}.
\end{eqnarray}

$\left(N_\mathrm{p}\right)_{\mathrm{max}}$ is the fundamental limit on the
torque exerted on an isotropic electric dipolar particle by a circularly
polarized plane wave. The same bound is attained for an isotropic magnetic dipolar particle. 
$N^{\mathrm{norm}}$ will be later used for normalization. Note that for an isotropic particle, it can be easily deduced from Eq.~\ref{eq:Np} that  a linearly polarized
plane wave exerts no torque $N=0$.

Figure~\ref{fig: ED_LossFactor} (b) shows the maximal
force and torque at resonance exerted by a plane wave as a function of the
loss factor, i.e. $\gamma_\mathrm{nr}/\gamma_\mathrm{r}$. In the over-coupling regime, $C_\mathrm{ext}$, and consequently the force is maximized. On the other hand, in the critical coupling regime, the  $C_\mathrm{abs}$ and consequently the torque is maximized.

\textit{Multipole expansion}: To extend our analysis to include multiple MCs, we go beyond the single channel dipole approximation. In the multipolar expansion, the
incident and scattered fields are expanded as~\cite{Jackson1999,Xu1995c,muhlig2011multipole}:
\begin{eqnarray}
\mathbf{E}_{\mathrm{i\mathrm{nc}}} & = & -\sum_{j=1}^{\infty}\sum_{m=-j}^{j}E_{jm}\left(p_{jm}\mathbf{N}_{jm}^{\left(1\right)}+\mathrm{i}q_{jm}\mathbf{M}_{jm}^{\left(1\right)}\right),\nonumber \\
\mathbf{E}_{\mathrm{sca}} & = & \sum_{j=1}^{\infty}\sum_{m=-j}^{j}E_{jm}\left(a_{jm}\mathbf{N}_{jm}^{\left(3\right)}+\mathrm{i}b_{jm}\mathbf{M}_{jm}^{\left(3\right)}\right),
\end{eqnarray}
\noindent where $\left[\mathbf{M}_{jm}^{\left(1\right)}\left(\mathbf{r};\omega\right),\mathbf{N}_{jm}^{\left(1\right)}\left(\mathbf{r};\omega\right)\right]$ are the regular 
and $\left[\mathbf{M}_{jm}^{\left(3\right)}\left(\mathbf{r};\omega\right),\mathbf{N}_{jm}^{\left(3\right)}\left(\mathbf{r};\omega\right)\right]$
the outgoing vector spherical wave functions (VSWFs).
$\left(p_{jm},q_{jm}\right)$ and $\left(a_{jm},b_{jm}\right)$ are
the amplitudes of the VSWFs expanding the incident and scattered fields. $j(j+1)$ is the eigenvalue of the AM squared $\mathbf{J}^2$, and $m$ is the eigenvalue of the z-component of the AM $\mathbf{J}_z$. $E_{jm}$ is a
normalizing factor (\textit{SM}).

The incident fields are assumed to be known. Their VSWF amplitudes
can be calculated using orthogonality relations. The VSWF amplitudes
of a plane wave are given in the
\textit{SM}. Finding the scattered field amplitudes is
not a trivial task. However, Mie theory provides analytical solutions for an isotropic particle ~\cite{mie1908beitrage,Xu1995c}.
The VSWF amplitudes of the scattered and incident fields are related by $a_{jm}=a_{j}p_{jm}$, and $b_{jm}=b_{j}q_{jm}$, with $a_{j}$ and $b_{j}$ being the electric and magnetic Mie coefficients \cite{Xu1995c}. 
Each Mie coefficient has a spectral profile with certain resonance peaks. For any of these channels (Mie channels) the angular momentum and parity of the fields are preserved and no energy cross-coupling occurs among different MCs. Based 
on energy conservation, the Mie coefficients are always smaller than unity and at resonance of a non-absorbing particle they are equal to unity ~\cite{tribelsky2006anomalous}.

\begin{table*}
\protect\caption{Fundamental limits on optical force constituents\label{tabel:Force}}

\resizebox{\textwidth}{!}{

\begin{centering}
\begin{tabular}{|c|c|c|c|c|}
\hline
Term & %
\begin{tabular}{c}
\vspace*{-0.3cm}
\tabularnewline
$F/F^{\mathrm{norm}}$\tabularnewline
\vspace*{-0.3cm}
\tabularnewline
\end{tabular} & $(F)_{\mathrm{max}}/F^{\mathrm{norm}}$ & Maximum at (over-coupling regime) & Example \tabularnewline
\hline
\hline
$F_{j^\mathrm{e}}$ & %
\begin{tabular}{c}
\vspace*{-0.3cm}
\tabularnewline
$\frac{2\pi}{\lambda^{2}}C_{\mathrm{ext},j^\mathrm{e}}=\left(2j+1\right)\Re\left(a_{j}\right)$\tabularnewline
\vspace*{-0.3cm}
\tabularnewline
\end{tabular} & $\left(2j+1\right)$ & $\left|a_{j}\right|$ resonance & dielectric sphere (Fig.~\ref{fig:Losslss})\tabularnewline
\hline
$F_{j^\mathrm{m}}$ & %
\begin{tabular}{c}
\vspace*{-0.3cm}
\tabularnewline
$\frac{2\pi}{\lambda^{2}}C_{\mathrm{ext,}j^\mathrm{m}}=\left(2j+1\right)\Re\left(b_{j}\right)$\tabularnewline
\vspace*{-0.3cm}
\tabularnewline
\end{tabular} & $\left(2j+1\right)$ & $\left|b_{j}\right|$ resonance & dielectric sphere (Fig.~\ref{fig:Losslss})\tabularnewline
\hline
$F_{j^\mathrm{e}j^\mathrm{m}}$ & %
\begin{tabular}{c}
\vspace*{-0.3cm}
\tabularnewline
$-\frac{2\left(2j+1\right)}{j\left(j+1\right)}\Re\left(a_{j}b_{j}^{*}\right)$\tabularnewline
\vspace*{-0.3cm}
\tabularnewline
\end{tabular} & $-\frac{2\left(2j+1\right)}{j\left(j+1\right)}$ & simultaneous $\left|a_{j}\right|$ and $\left|b_{j}\right|$ resonance & dual dielectric sphere (Fig.~S3)\tabularnewline
\hline
$F_{j^\mathrm{e}\left(j+1\right)^{e}}$ & %
\begin{tabular}{c}
\vspace*{-0.3cm}
\tabularnewline
$-\frac{2j\left(j+2\right)}{\left(j+1\right)}\Re\left(a_{j+1}a_{j}^{*}\right)$\tabularnewline
\vspace*{-0.3cm}
\tabularnewline
\end{tabular} & $-\frac{2j\left(j+2\right)}{\left(j+1\right)}$ & simultaneous $\left|a_{j}\right|$ and $\left|a_{j+1}\right|$ resonance & dielectric core-multishell (Fig.~S4)\tabularnewline
\hline
$F_{j^\mathrm{m}\left(j+1\right)^{m}}$ & %
\begin{tabular}{c}
\vspace*{-0.3cm}
\tabularnewline
$-\frac{2j\left(j+2\right)}{\left(j+1\right)}\Re\left(b_{j+1}b_{j}^{*}\right)$\tabularnewline
\vspace*{-0.3cm}
\tabularnewline
\end{tabular} & $-\frac{2j\left(j+2\right)}{\left(j+1\right)}$ & simultaneous $\left|b_{j}\right|$ and $\left|b_{j+1}\right|$ resonance & dielectric core-multishell\tabularnewline
\hline
\end{tabular}
\par\end{centering}

}
\end{table*}
\begin{table*}
\protect\caption{Fundamental limits on optical torque constituents\label{tabel:Torque}}

\resizebox{\textwidth}{!}{

\begin{centering}
\begin{tabular}{|c|c|c|c|c|}
\hline
Term & %
\begin{tabular}{c}
\vspace*{-0.3cm}
\tabularnewline
$N/N^{\mathrm{norm}}$\tabularnewline
\vspace*{-0.3cm}
\tabularnewline
\end{tabular} & $(N)_{\mathrm{max}}/N^{\mathrm{norm}}$ & Maximum at (critical-coupling regime) & Example \tabularnewline
\hline
\hline
$N_{j^\mathrm{e}}$ & %
\begin{tabular}{c}
\vspace*{-0.3cm}
\tabularnewline
$\frac{\sigma I_{0}}{\omega}C_{\mathrm{abs},j^\mathrm{e}}=4\sigma\left(2j+1\right)\Re\left(a_{j}-\left|a_{j}\right|^{2}\right)$\tabularnewline
\vspace*{-0.3cm}
\tabularnewline
\end{tabular} & $\sigma\left(2j+1\right)$ & $\left|a_{j}\right|$ resonance & dielectric sphere (Fig.~S5)\tabularnewline
\hline
$N_{j^\mathrm{m}}$ & %
\begin{tabular}{c}
\vspace*{-0.3cm}
\tabularnewline
$\frac{\sigma I_{0}}{\omega}C_{\mathrm{abs},j^\mathrm{m}}=4\sigma\left(2j+1\right)\Re\left(b_{j}-\left|b_{j}\right|^{2}\right)$\tabularnewline
\vspace*{-0.3cm}
\tabularnewline
\end{tabular} & $\sigma\left(2j+1\right)$ & $\left|b_{j}\right|$ resonance & dielectric sphere (Figs.~\ref{fig:Absorptive_dielectric_sphere:}-S5)\tabularnewline
\hline
\end{tabular}
\par\end{centering}
}
\end{table*}

\textit{Fundamental limits on optical force}:
The multipolar description of the force on a
particle by an arbitrary illumination has been presented in~\cite{barton1989theoretical,almaas1995radiation}
(\textit{SM}). Considering the contributions of
different MCs, the force on an isotropic particle in along $+z$ is derived as:
\noindent
\begin{eqnarray}
F & = & \sum_{j=1}^{\infty}\left\{ F_{j^\mathrm{e}}+F_{j^\mathrm{m}}+F_{j^\mathrm{e}j^\mathrm{m}}+F_{j^\mathrm{e}\left(j+1\right)^\mathrm{e}}+F_{j^\mathrm{m}\left(j+1\right)^\mathrm{m}}\right\} \nonumber \\
 & = & \left[F_\mathrm{p}+F_\mathrm{Q^{e}}+F_\mathrm{O^{e}}+...\right]+\left[F_\mathrm{m}+F_\mathrm{Q^{m}}+F_\mathrm{O^{m}}+...\right]\nonumber \\
 & + & \left[F_\mathrm{pm}+F_\mathrm{Q^{e}Q^{m}}+F_\mathrm{O^{e}O^{m}}+...\right]\nonumber \\
 & + & \left[F_\mathrm{pQ^{e}}+F_\mathrm{Q^{e}O^{e}}+...\right]+\left[F_\mathrm{mQ^{m}}+F_\mathrm{Q^{m}O^{m}}+...\right].\label{eq:force-n-1}
\end{eqnarray}

 $F_{j^\mathrm{e}}$ ($F_{j^\mathrm{m}}$) is the force due to an individual electric (magnetic) MC.
$F_{j^\mathrm{e}j^\mathrm{m}}$ is the force due to the spectral interference of two MCs with identical $j$ but opposite character. $F_{j^\mathrm{e}\left(j+1\right)^\mathrm{e}}$ ($F_{j^\mathrm{m}\left(j+1\right)^\mathrm{m}}$) 
is due to the spectral interference of two electric (magnetic) MCs with the same character and $j$ and $j+1$ total AM number.

\begin{figure}
\begin{centering}
\includegraphics[scale=0.36]{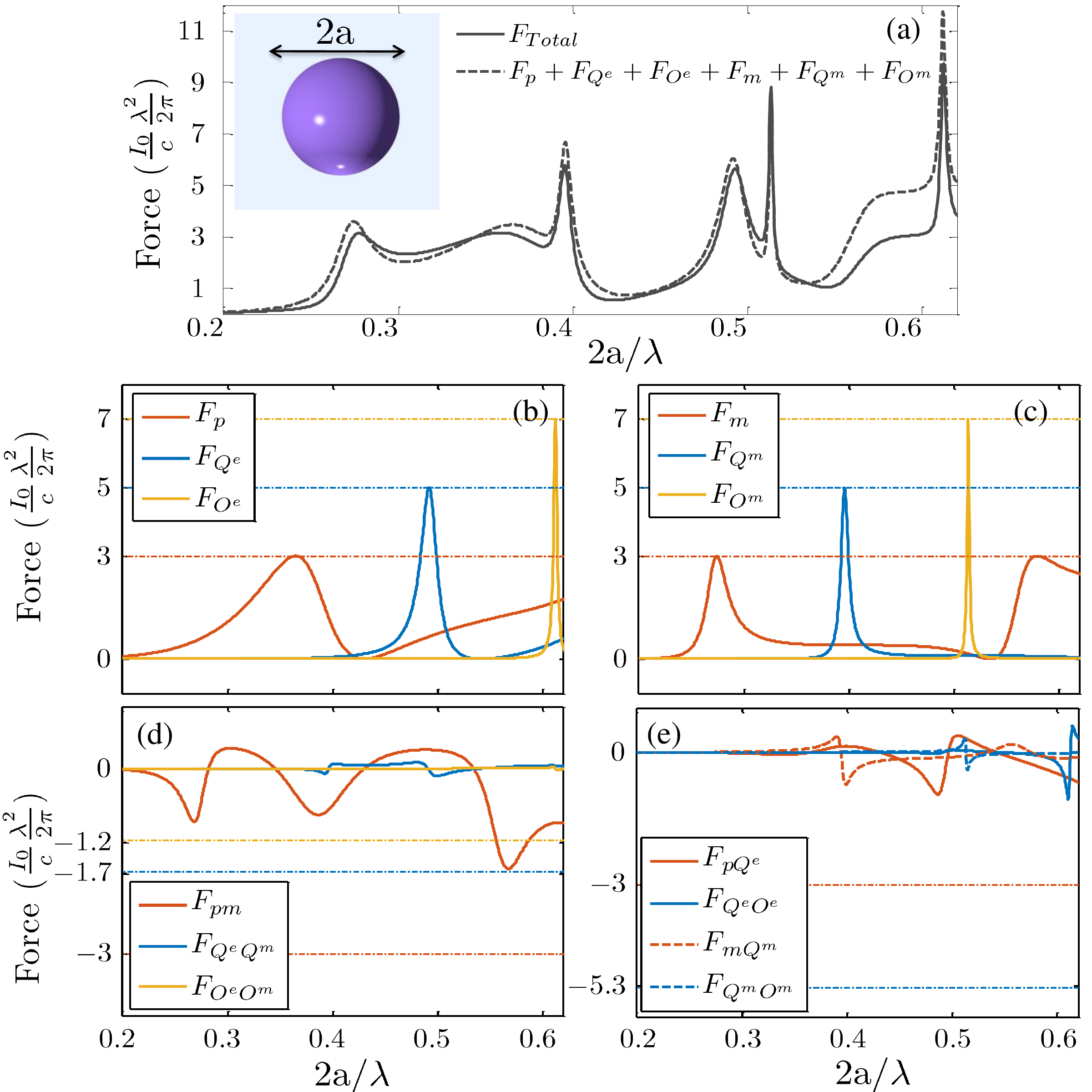}
\par\end{centering}

\protect\caption{\textit{Non-absorbing dielectric sphere}: (a) Optical force exerted by an arbitrarily polarized plane wave on a non-absorbing sphere ($\epsilon_\mathrm{r} = (3.5)^2$, $\mu_\mathrm{r}=1$) depending on
the sphere's size parameter (solid line). Contributions of the non-interference terms (dashed line). (b)
The individual contributions of the dipole, quadrupole, and octopole electric and (c) magnetic
MCs. (d) The partial contribution of the interference of dipole-dipole, quadrupole-quadrupole
and octopole-octopole electric and magnetic MCs. (e) The interference of dipole-quadrupole and quadrupole-octopole,
electric and magnetic MCs.\label{fig:Losslss}}
\end{figure}

\begin{figure}
\begin{centering}
\includegraphics[scale=0.4]{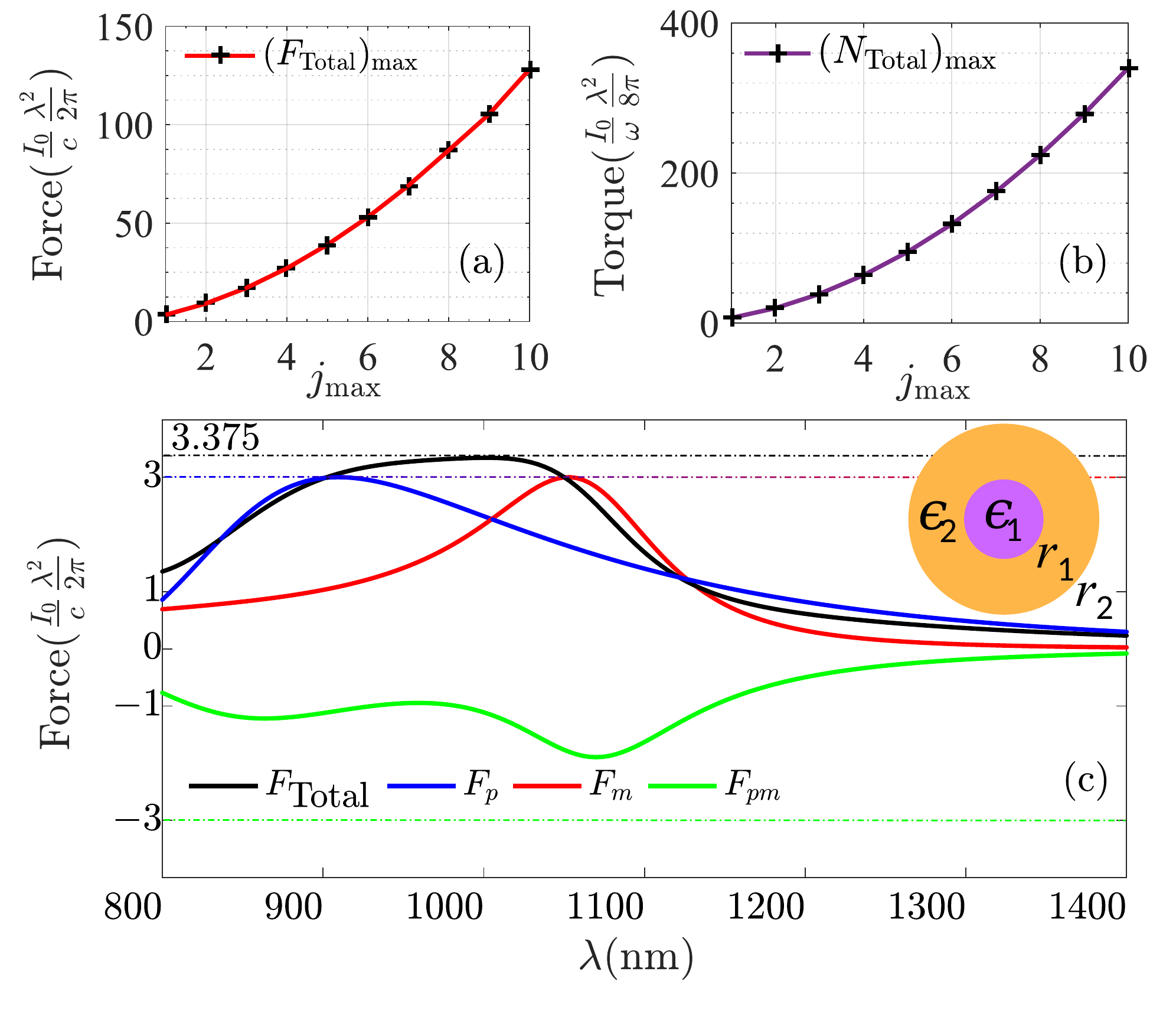}
\par\end{centering}
\protect\caption{ (a) Maximum total optical force and (b) torque as a function of the maximum non-negligible multipole moment order $j_\mathrm{max}$.  (c) Optical force exerted on an optimized isotropic 
non-absorbing core-shell particle, illuminated by an arbitrarily polarized plane wave. The figure also illustrates the contributions of the individual dipolar electric and magnetic  MCs and their interference. 
Parameters of the particle: $r_1$=88~nm, $r_2$=179~nm, $\epsilon_1=4^2$ and  $\epsilon_2=(2.64)^2$. \label{fig:Ftotal}}
\end{figure}

\begin{figure}
\begin{centering}
\includegraphics[scale=0.38]{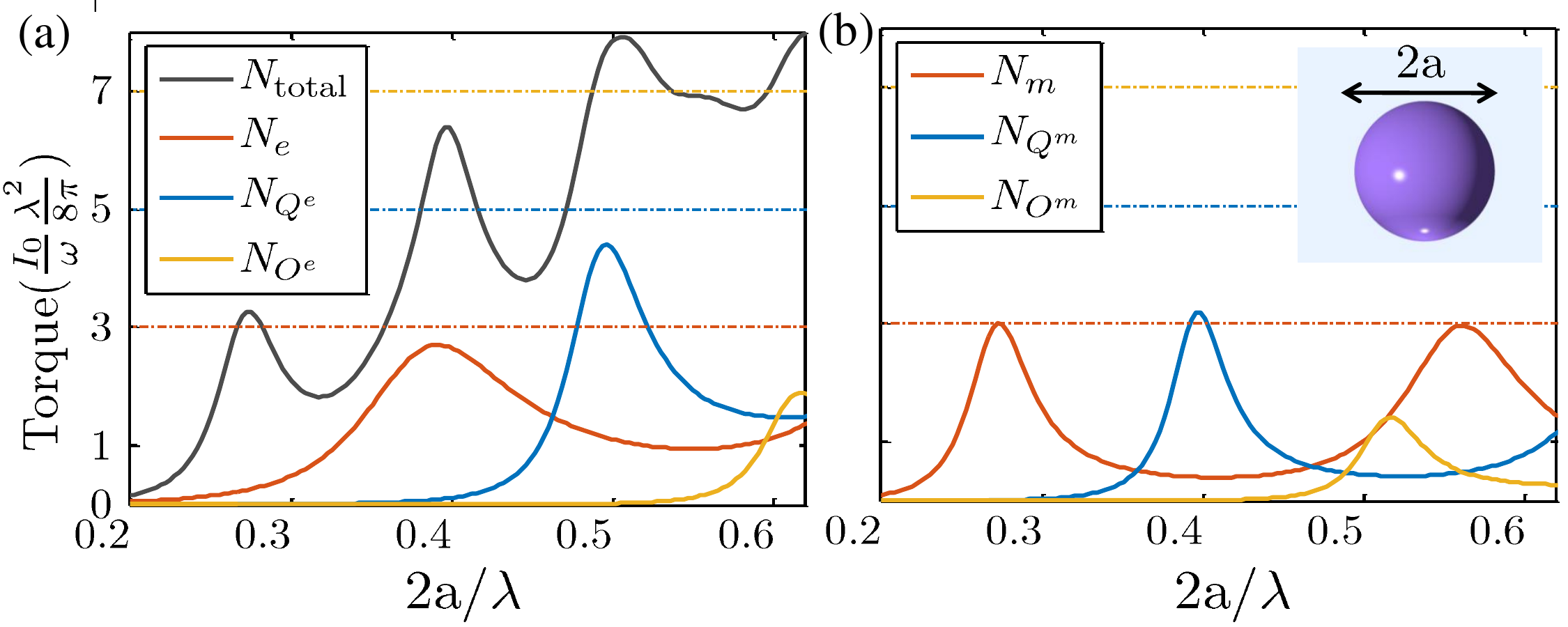}
\par\end{centering}

\protect\caption{\textit{Absorbing dielectric sphere}: (a) Optical torque on an absorbing sphere [$\epsilon_r = (3.5 +0.16i)^2$, $\mu_r=1$], by a circularly polarized plane wave as a function of the sphere's
size parameter. The individual contribution of the dipole, quadrupole,
and octopole electric and (b) magnetic MCs. \label{fig:Absorptive_dielectric_sphere:}}

\end{figure}

Assuming an arbitrarily polarized plane wave, propagating in +z direction and illuminating an isotropic particle, we have derived the expression for different
force terms and the conditions to maximize their individual contribution (\textit{SM}). The results are shown in Table~\ref{tabel:Force}. Specific examples are mentioned that maximize each term. Figure~\ref{fig:Losslss} shows different non-zero contributions
to the optical force exerted on a non-absorbing dielectric sphere up to the multipole order $j=3$ (In Fig.~S2 the same is considered for an absorbing particle, which shows a significant damping in the force near resonance). 
For all the upcoming figures, the maximum contribution of each force term is shown by a same color dashed line. Let us now focus on maximizing the total optical force. For a homogeneous sphere like the one in Fig.~\ref{fig:Losslss}, at least for the lower size parameter values, where the spectral overlap of MCs is small, the interference terms are small and the optical force can be approximated by the 
individual contribution of MCs. Therefore, in this case, the maximal total optical force is well approximated by the maximum of the individual contribution of the MCs, $(2j+1)F^{\mathrm{norm}}$. 
The individual contribution of a MC is directly related to the extinction cross section (SM). Although the $(2j+1)$ factor is bigger for higher multipoles, 
the resonance wavelengths of the channels are lower and hence in general, for an isotropic particle, the achievable maximum force is smaller for higher order multipoles.  

When extending our analysis to more general isotropic particles, interference terms come into play. Figure~S3
considers the optical force on a dual ~\cite{Fernandez-Corbaton2013c} dielectric sphere.
Due to the simultaneous resonance of the electric ($a_{j}$) and magnetic
($b_{j}$) MCs, the contribution of the interference term $F_{j^\mathrm{e}j^\mathrm{m}}$ is maximized. In Fig.~S4, a core-multishell particle is analysed, where the electric dipole-quadrupole 
and quadrupole-octopole interference force contributions are maximized $(F_\mathrm{pQ^e}=-3F^{\mathrm{norm}},F_\mathrm{Q^eO^e}\approx-5.3F^{\mathrm{norm}})$. Note that the maximum contribution of the 
interference terms is always negative. However, as proven in SM, the maximum positive contribution  of  an interference term (at off-resonance) is 8 times smaller than the maximum negative contribution (at resonance). 

To calculate the maximum total optical force, the contribution of interference terms should be considered. For an isotropic particle with finite volume and finite permeability and permittivity (positive or negative), which is illuminated by a plane wave at a given frequency, avoiding pathological cases, the optical response to any order of accuracy is assumed to be describable with a finite number of multipole moments and hence the total optical force is always finite. For a non-absorbing particle, each Mie coefficient 
is modeled by a simple formula ($a_j=\cos\alpha_j\exp\mathrm{i}\alpha_j$, $b_j=\cos\beta_j\exp\mathrm{i}\beta_j$) with a single real-valued variable $\alpha_j$ or $\beta_j$ \cite{hulst1957light}. Based on this model and Eq.~\ref{eq:force-n-1}, assuming that the resonance of the
MCs can be optimally engineered (i.e. assuming all $\alpha_j$ and $\beta_j$s to be independent), the maximum force can be calculated as a function of the maximum  non-negligible multipole 
order $j_\mathrm{max}$. For a smaller number of Mie channels the maximal force can be derived analytically, i.e. up to $j_\mathrm{max}=3$. For a larger number of Mie channels a genetic algorithm has been used (SM). The results are shown in Fig.~\ref{fig:Ftotal}(a). For a dipolar particle, the maximum of the optical 
force is $3.375F^{\mathrm{norm}}$. This upper bound for the force is not met at the resonance due to the interference contribution. Using particle swarm optimization (PSO)~\cite{lee2008modern}, we have optimized a dielectric core-multishell particle
that receives the maximum optical force in dipolar approximation at $\lambda=1$ $\mu$m [Fig.~\ref{fig:Ftotal}(c)] (SM). This demonstrates the applicability of the present formalism.

\textit{Fundamental limits on optical torque}:
In a similar approach, the z-component of the optical torque exerted on an isotropic
particle by an arbitrary illumination, presented in Ref.~\onlinecite{barton1989theoretical}, can be rewritten as:
\begin{eqnarray}
N & = & \sum_{j=1}^{\infty}\left\{ N_{j^\mathrm{e}}+N_{j^\mathrm{m}}\right\} =\left[N_\mathrm{p}+N_\mathrm{Q^{e}}+N_\mathrm{O^{e}}+...\right]\nonumber\\
 & + & \left[N_\mathrm{m}+N_\mathrm{Q^{m}}+N_\mathrm{O^{m}}+...\right],
\end{eqnarray}
\noindent where $N_{\mathrm{j^\mathrm{e}}}$ ($N_{\mathrm{j^\mathrm{m}}}$) is
the contribution of an electric (magnetic)
MC to the optical torque. The contribution of an electric MC to the torque is:
\begin{eqnarray}
N_{j^\mathrm{e}} =\frac{\lambda^{3}I_{0}}{8\pi^{3}c}\Re\left(a_{j}-\left|a_{j}\right|^{2}\right)\sum_{m=-j}^{j}m\left|p_{jm}\right|^{2}=\frac{\sigma I_{0}C_{\mathrm{abs},j^\mathrm{e}}}{\omega}.
\end{eqnarray}
The exerted torque is directly 
related to the averaged total AM~\cite{Jackson1999} of the incident wave in the z-direction.
 For a circularly polarized plane wave propagating along z, the contribution of an electric MC to the torque simplifies to:
\begin{eqnarray}
N_{j^\mathrm{e}} & = & 4\sigma\left(2j+1\right)\Re\left(a_{j}-\left|a_{j}\right|^{2}\right)N^{\mathrm{norm}}.
\end{eqnarray}

For an absorbing isotropic particle at the critical coupling regime, $\Re\left(a_{j}-\left|a_{j}\right|^{2}\right)$ is maximum and equal to $0.25$
\cite{miroshnichenko2016ultimate}. Therefore, the maximum torque is derived as:
\begin{eqnarray}
(N_{j^\mathrm{e}})_{\mathrm{max}} & = & \sigma\left(2j+1\right)N^{\mathrm{norm}}.
\end{eqnarray}

A similar relation can be derived for the  contribution of a magnetic MC. Table~\ref{tabel:Torque} summarizes the two contributions and the conditions for maximizing them. To maximize the total optical torque, $C_\mathrm{abs}$ should be maximized, i.e. the critical coupling condition should be satisfied. The critical coupling is met at a single frequency. In Fig.~\ref{fig:Absorptive_dielectric_sphere:},
the optical torque is critically coupled at the magnetic dipole MC resonance.

Unlike the force, the torque along +z, does not have interference terms and contribution of MCs directly add up. Therefore, for a dual (identical $a_j$ and $b_j$) sphere that is critically coupled, 
from Table~\ref{tabel:Torque}, the torque is $(N_{j^\mathrm{e}})_{\mathrm{max}}+(N_{j^\mathrm{m}})_{\mathrm{max}}=2(2j+1)N^{\mathrm{norm}}$. This is shown in Fig.~S5. As a simple extrapolation, it can be concluded that the maximum 
optical torque on a critically coupled particle occurs when all the multipole moments overlap resonantly. Therefore, the maximum optical torque as a
function of $j_\mathrm{max}$ is $\sum_{j=1}^{j_\mathrm{max}}2(2j+1)N^{\mathrm{norm}}=2j_\mathrm{max}(j_\mathrm{max}+2)$. This relation is plotted in Fig.~\ref{fig:Ftotal}(b).

For completeness, we have calculated the optical force and torque on a silver nanosphere (Fig.~S6) to compare our results with a realistic absorbing particle. Our results can be used to design super-accelerable and -rotatable particles by engineering the spectral resonance of the MCs. The designed 
particles can in turn be used in opto-nanorobots.

\section*{Acknowledgements}
We thank Hadi Rahmanpanah and Ahvand Jalali for their help in the optimizations. We acknowledge partial financial support by the Deutsche Forschungsgemeinschaft through CRC 1173. A.R. also acknowledges support from the Karlsruhe School of Optics and Photonics (KSOP).


\begin{thebibliography}{36}
\expandafter\ifx\csname natexlab\endcsname\relax\def\natexlab#1{#1}\fi
\expandafter\ifx\csname bibnamefont\endcsname\relax
  \def\bibnamefont#1{#1}\fi
\expandafter\ifx\csname bibfnamefont\endcsname\relax
  \def\bibfnamefont#1{#1}\fi
\expandafter\ifx\csname citenamefont\endcsname\relax
  \def\citenamefont#1{#1}\fi
\expandafter\ifx\csname url\endcsname\relax
  \def\url#1{\texttt{#1}}\fi
\expandafter\ifx\csname urlprefix\endcsname\relax\def\urlprefix{URL }\fi
\providecommand{\bibinfo}[2]{#2}
\providecommand{\eprint}[2][]{\url{#2}}

\bibitem[{\citenamefont{Jackson}(1999)}]{Jackson1999}
\bibinfo{author}{\bibfnamefont{J.~D.} \bibnamefont{Jackson}},
  \emph{\bibinfo{title}{{Classical Electrodynamics}}}
  (\bibinfo{publisher}{Wiley}, \bibinfo{year}{1999}).

\bibitem[{\citenamefont{Bohren and Huffman}(2008)}]{Bohren2008}
\bibinfo{author}{\bibfnamefont{C.~F.} \bibnamefont{Bohren}} \bibnamefont{and}
  \bibinfo{author}{\bibfnamefont{D.~R.} \bibnamefont{Huffman}},
  \emph{\bibinfo{title}{Absorption and Scattering of Light by Small Particles}}
  (\bibinfo{publisher}{John Wiley {\&} Sons}, \bibinfo{year}{2008}).

\bibitem[{\citenamefont{Xu}(1995)}]{Xu1995c}
\bibinfo{author}{\bibfnamefont{Y.}~\bibnamefont{Xu}}, \bibinfo{journal}{App.
  Opt.} \textbf{\bibinfo{volume}{34}}, \bibinfo{pages}{4573}
  (\bibinfo{year}{1995}).

\bibitem[{\citenamefont{Ruan and Fan}(2010)}]{ruan2010superscattering}
\bibinfo{author}{\bibfnamefont{Z.}~\bibnamefont{Ruan}} \bibnamefont{and}
  \bibinfo{author}{\bibfnamefont{S.}~\bibnamefont{Fan}},
  \bibinfo{journal}{Phys. Rev. Lett.} \textbf{\bibinfo{volume}{105}},
  \bibinfo{pages}{013901} (\bibinfo{year}{2010}).

\bibitem[{\citenamefont{Ruan and Fan}(2011)}]{Ruan2011a}
\bibinfo{author}{\bibfnamefont{Z.}~\bibnamefont{Ruan}} \bibnamefont{and}
  \bibinfo{author}{\bibfnamefont{S.}~\bibnamefont{Fan}}, \bibinfo{journal}{App.
  Phys. Lett.} \textbf{\bibinfo{volume}{98}}, \bibinfo{pages}{043101}
  (\bibinfo{year}{2011}).

\bibitem[{\citenamefont{Miroshnichenko and
  Tribelsky}(2016)}]{miroshnichenko2016ultimate}
\bibinfo{author}{\bibfnamefont{A.~E.} \bibnamefont{Miroshnichenko}}
  \bibnamefont{and} \bibinfo{author}{\bibfnamefont{M.~I.}
  \bibnamefont{Tribelsky}}, \bibinfo{journal}{arXiv preprint arXiv:1603.03513}
  (\bibinfo{year}{2016}).

\bibitem[{\citenamefont{Maslovski et~al.}(2016)\citenamefont{Maslovski,
  Simovski, and Tretyakov}}]{maslovski2016overcoming}
\bibinfo{author}{\bibfnamefont{S.~I.} \bibnamefont{Maslovski}},
  \bibinfo{author}{\bibfnamefont{C.~R.} \bibnamefont{Simovski}},
  \bibnamefont{and} \bibinfo{author}{\bibfnamefont{S.~A.}
  \bibnamefont{Tretyakov}}, \bibinfo{journal}{New J. Phys.}
  \textbf{\bibinfo{volume}{18}}, \bibinfo{pages}{013034}
  (\bibinfo{year}{2016}).

\bibitem[{\citenamefont{Cox et~al.}(2002)\citenamefont{Cox, DeWeerd, and
  Linden}}]{cox2002experiment}
\bibinfo{author}{\bibfnamefont{A.}~\bibnamefont{Cox}},
  \bibinfo{author}{\bibfnamefont{A.~J.} \bibnamefont{DeWeerd}},
  \bibnamefont{and} \bibinfo{author}{\bibfnamefont{J.}~\bibnamefont{Linden}},
  \bibinfo{journal}{Am. J. Phys.} \textbf{\bibinfo{volume}{70}},
  \bibinfo{pages}{620} (\bibinfo{year}{2002}).

\bibitem[{\citenamefont{Yang et~al.}(2008)\citenamefont{Yang, Chen, Luo, and
  Ma}}]{yang2008superscatterer}
\bibinfo{author}{\bibfnamefont{T.}~\bibnamefont{Yang}},
  \bibinfo{author}{\bibfnamefont{H.}~\bibnamefont{Chen}},
  \bibinfo{author}{\bibfnamefont{X.}~\bibnamefont{Luo}}, \bibnamefont{and}
  \bibinfo{author}{\bibfnamefont{H.}~\bibnamefont{Ma}},
  \bibinfo{journal}{Optics express} \textbf{\bibinfo{volume}{16}},
  \bibinfo{pages}{18545} (\bibinfo{year}{2008}).

\bibitem[{\citenamefont{Chen et~al.}(2011)\citenamefont{Chen, Ng, Lin, and
  Chan}}]{chen2011optical}
\bibinfo{author}{\bibfnamefont{J.}~\bibnamefont{Chen}},
  \bibinfo{author}{\bibfnamefont{J.}~\bibnamefont{Ng}},
  \bibinfo{author}{\bibfnamefont{Z.}~\bibnamefont{Lin}}, \bibnamefont{and}
  \bibinfo{author}{\bibfnamefont{C.}~\bibnamefont{Chan}},
  \bibinfo{journal}{Nat. Photonics} \textbf{\bibinfo{volume}{5}},
  \bibinfo{pages}{531} (\bibinfo{year}{2011}).

\bibitem[{\citenamefont{Nieto-Vesperinas}(2015)}]{nieto2015optical}
\bibinfo{author}{\bibfnamefont{M.}~\bibnamefont{Nieto-Vesperinas}},
  \bibinfo{journal}{Opt. Lett.} \textbf{\bibinfo{volume}{40}},
  \bibinfo{pages}{3021} (\bibinfo{year}{2015}).

\bibitem[{\citenamefont{Jiang et~al.}(2015)\citenamefont{Jiang, Chen, Chen, Ng,
  and Lin}}]{jiang2015poynting}
\bibinfo{author}{\bibfnamefont{Y.}~\bibnamefont{Jiang}},
  \bibinfo{author}{\bibfnamefont{H.}~\bibnamefont{Chen}},
  \bibinfo{author}{\bibfnamefont{J.}~\bibnamefont{Chen}},
  \bibinfo{author}{\bibfnamefont{J.}~\bibnamefont{Ng}}, \bibnamefont{and}
  \bibinfo{author}{\bibfnamefont{Z.}~\bibnamefont{Lin}},
  \bibinfo{journal}{arXiv preprint arXiv:1511.08546}  (\bibinfo{year}{2015}).

\bibitem[{\citenamefont{{Li Yangcheng} et~al.}(2013)\citenamefont{{Li
  Yangcheng}, {Svitelskiy Oleksiy V}, {Maslov Alexey V}, {Carnegie David},
  {Rafailov Edik}, and {Astratov Vasily N}}}]{Yangcheng2013giant}
\bibinfo{author}{\bibnamefont{{Li Yangcheng}}},
  \bibinfo{author}{\bibnamefont{{Svitelskiy Oleksiy V}}},
  \bibinfo{author}{\bibnamefont{{Maslov Alexey V}}},
  \bibinfo{author}{\bibnamefont{{Carnegie David}}},
  \bibinfo{author}{\bibnamefont{{Rafailov Edik}}}, \bibnamefont{and}
  \bibinfo{author}{\bibnamefont{{Astratov Vasily N}}}, \bibinfo{journal}{Light
  Sci Appl} \textbf{\bibinfo{volume}{2}}, \bibinfo{pages}{e64}
  (\bibinfo{year}{2013}).

\bibitem[{\citenamefont{Dienerowitz et~al.}(2008)\citenamefont{Dienerowitz,
  Mazilu, and Dholakia}}]{dienerowitz2008optical}
\bibinfo{author}{\bibfnamefont{M.}~\bibnamefont{Dienerowitz}},
  \bibinfo{author}{\bibfnamefont{M.}~\bibnamefont{Mazilu}}, \bibnamefont{and}
  \bibinfo{author}{\bibfnamefont{K.}~\bibnamefont{Dholakia}},
  \bibinfo{journal}{J. Nanophotonics} \textbf{\bibinfo{volume}{2}},
  \bibinfo{pages}{021875} (\bibinfo{year}{2008}).

\bibitem[{\citenamefont{Grier}(2003)}]{Grier2003}
\bibinfo{author}{\bibfnamefont{D.~G.} \bibnamefont{Grier}},
  \bibinfo{journal}{Nature} \textbf{\bibinfo{volume}{424}},
  \bibinfo{pages}{810} (\bibinfo{year}{2003}).

\bibitem[{\citenamefont{Yannopapas}(2008)}]{yannopapas2008optical}
\bibinfo{author}{\bibfnamefont{V.}~\bibnamefont{Yannopapas}},
  \bibinfo{journal}{Phys. Rev. B} \textbf{\bibinfo{volume}{78}},
  \bibinfo{pages}{045412} (\bibinfo{year}{2008}).

\bibitem[{\citenamefont{Rahimzadegan et~al.}(2016)\citenamefont{Rahimzadegan,
  Fruhnert, Alaee, Fernandez-Corbaton, and
  Rockstuhl}}]{rahimzadegan2016optical}
\bibinfo{author}{\bibfnamefont{A.}~\bibnamefont{Rahimzadegan}},
  \bibinfo{author}{\bibfnamefont{M.}~\bibnamefont{Fruhnert}},
  \bibinfo{author}{\bibfnamefont{R.}~\bibnamefont{Alaee}},
  \bibinfo{author}{\bibfnamefont{I.}~\bibnamefont{Fernandez-Corbaton}},
  \bibnamefont{and}
  \bibinfo{author}{\bibfnamefont{C.}~\bibnamefont{Rockstuhl}},
  \bibinfo{journal}{Phys. Rev. B} \textbf{\bibinfo{volume}{94}},
  \bibinfo{pages}{125123} (\bibinfo{year}{2016}).

\bibitem[{\citenamefont{{\v{C}}i{\v{z}}m{\'a}r
  et~al.}(2006)\citenamefont{{\v{C}}i{\v{z}}m{\'a}r, {\v{S}}iler,
  {\v{S}}er{\`y}, Zem{\'a}nek, Garc{\'e}s-Ch{\'a}vez, and
  Dholakia}}]{vcivzmar2006optical}
\bibinfo{author}{\bibfnamefont{T.}~\bibnamefont{{\v{C}}i{\v{z}}m{\'a}r}},
  \bibinfo{author}{\bibfnamefont{M.}~\bibnamefont{{\v{S}}iler}},
  \bibinfo{author}{\bibfnamefont{M.}~\bibnamefont{{\v{S}}er{\`y}}},
  \bibinfo{author}{\bibfnamefont{P.}~\bibnamefont{Zem{\'a}nek}},
  \bibinfo{author}{\bibfnamefont{V.}~\bibnamefont{Garc{\'e}s-Ch{\'a}vez}},
  \bibnamefont{and} \bibinfo{author}{\bibfnamefont{K.}~\bibnamefont{Dholakia}},
  \bibinfo{journal}{Phys. Rev. B} \textbf{\bibinfo{volume}{74}},
  \bibinfo{pages}{035105} (\bibinfo{year}{2006}).

\bibitem[{\citenamefont{Mavroidis and
  Ferreira}(2013)}]{mavroidis2013nanorobotics}
\bibinfo{author}{\bibfnamefont{C.}~\bibnamefont{Mavroidis}} \bibnamefont{and}
  \bibinfo{author}{\bibfnamefont{A.}~\bibnamefont{Ferreira}}, in
  \emph{\bibinfo{booktitle}{Nanorobotics}} (\bibinfo{publisher}{Springer},
  \bibinfo{year}{2013}), pp. \bibinfo{pages}{3--27}.

\bibitem[{\citenamefont{Nieminen et~al.}(2005)\citenamefont{Nieminen, Higuet,
  Kn{\"o}ner, Loke, Parkin, Singer, Heckenberg, and
  Rubinsztein-Dunlop}}]{nieminen2005optically}
\bibinfo{author}{\bibfnamefont{T.~A.} \bibnamefont{Nieminen}},
  \bibinfo{author}{\bibfnamefont{J.}~\bibnamefont{Higuet}},
  \bibinfo{author}{\bibfnamefont{G.~G.} \bibnamefont{Kn{\"o}ner}},
  \bibinfo{author}{\bibfnamefont{V.~L.} \bibnamefont{Loke}},
  \bibinfo{author}{\bibfnamefont{S.}~\bibnamefont{Parkin}},
  \bibinfo{author}{\bibfnamefont{W.}~\bibnamefont{Singer}},
  \bibinfo{author}{\bibfnamefont{N.~R.} \bibnamefont{Heckenberg}},
  \bibnamefont{and}
  \bibinfo{author}{\bibfnamefont{H.}~\bibnamefont{Rubinsztein-Dunlop}}, in
  \emph{\bibinfo{booktitle}{Microelectronics, MEMS, and Nanotechnology}}
  (\bibinfo{organization}{International Society for Optics and Photonics},
  \bibinfo{year}{2005}), pp. \bibinfo{pages}{603813--603813}.

\bibitem[{\citenamefont{Stewart}(2005)}]{stewart2005angular}
\bibinfo{author}{\bibfnamefont{A.}~\bibnamefont{Stewart}}, \bibinfo{journal}{J.
  Mod. Opt.} \textbf{\bibinfo{volume}{52}}, \bibinfo{pages}{1145}
  (\bibinfo{year}{2005}).

\bibitem[{\citenamefont{Franke-Arnold et~al.}(2008)\citenamefont{Franke-Arnold,
  Allen, and Padgett}}]{franke2008advances}
\bibinfo{author}{\bibfnamefont{S.}~\bibnamefont{Franke-Arnold}},
  \bibinfo{author}{\bibfnamefont{L.}~\bibnamefont{Allen}}, \bibnamefont{and}
  \bibinfo{author}{\bibfnamefont{M.}~\bibnamefont{Padgett}},
  \bibinfo{journal}{Laser Photon. Rev.} \textbf{\bibinfo{volume}{2}},
  \bibinfo{pages}{299} (\bibinfo{year}{2008}).

\bibitem[{\citenamefont{Barnett}(2010)}]{barnett2010rotation}
\bibinfo{author}{\bibfnamefont{S.~M.} \bibnamefont{Barnett}},
  \bibinfo{journal}{J. Mod. Opt.} \textbf{\bibinfo{volume}{57}},
  \bibinfo{pages}{1339} (\bibinfo{year}{2010}).

\bibitem[{\citenamefont{Fernandez-Corbaton
  et~al.}(2012)\citenamefont{Fernandez-Corbaton, Zambrana-Puyalto, and
  Molina-Terriza}}]{Fernandez-Corbaton2012}
\bibinfo{author}{\bibfnamefont{I.}~\bibnamefont{Fernandez-Corbaton}},
  \bibinfo{author}{\bibfnamefont{X.}~\bibnamefont{Zambrana-Puyalto}},
  \bibnamefont{and}
  \bibinfo{author}{\bibfnamefont{G.}~\bibnamefont{Molina-Terriza}},
  \bibinfo{journal}{Phys. Rev. A} \textbf{\bibinfo{volume}{86}},
  \bibinfo{pages}{1} (\bibinfo{year}{2012}).

\bibitem[{\citenamefont{Long}(2011)}]{long2011deep}
\bibinfo{author}{\bibfnamefont{K.~F.} \bibnamefont{Long}},
  \emph{\bibinfo{title}{Deep Space Propulsion: A Roadmap to Interstellar
  Flight}} (\bibinfo{publisher}{Springer Science \& Business Media},
  \bibinfo{year}{2011}).

\bibitem[{\citenamefont{Nieto-Vesperinas
  et~al.}(2010)\citenamefont{Nieto-Vesperinas, S{\'a}enz, G{\'o}mez-Medina, and
  Chantada}}]{nieto2010optical}
\bibinfo{author}{\bibfnamefont{M.}~\bibnamefont{Nieto-Vesperinas}},
  \bibinfo{author}{\bibfnamefont{J.}~\bibnamefont{S{\'a}enz}},
  \bibinfo{author}{\bibfnamefont{R.}~\bibnamefont{G{\'o}mez-Medina}},
  \bibnamefont{and} \bibinfo{author}{\bibfnamefont{L.}~\bibnamefont{Chantada}},
  \bibinfo{journal}{Opt. Express} \textbf{\bibinfo{volume}{18}},
  \bibinfo{pages}{11428} (\bibinfo{year}{2010}).

\bibitem[{\citenamefont{Tretyakov}(2003)}]{tretyakov2003analytical}
\bibinfo{author}{\bibfnamefont{S.}~\bibnamefont{Tretyakov}},
  \emph{\bibinfo{title}{Analytical Modeling in Applied Electromagnetics}}
  (\bibinfo{publisher}{Artech House}, \bibinfo{year}{2003}).

\bibitem[{\citenamefont{Sipe and Kranendonk}(1974)}]{sipe1974macroscopic}
\bibinfo{author}{\bibfnamefont{J.}~\bibnamefont{Sipe}} \bibnamefont{and}
  \bibinfo{author}{\bibfnamefont{J.}~\bibnamefont{Kranendonk}},
  \bibinfo{journal}{Phys. Rev. A} \textbf{\bibinfo{volume}{9}},
  \bibinfo{pages}{1806} (\bibinfo{year}{1974}).

\bibitem[{\citenamefont{M{\"u}hlig et~al.}(2011)\citenamefont{M{\"u}hlig,
  Menzel, Rockstuhl, and Lederer}}]{muhlig2011multipole}
\bibinfo{author}{\bibfnamefont{S.}~\bibnamefont{M{\"u}hlig}},
  \bibinfo{author}{\bibfnamefont{C.}~\bibnamefont{Menzel}},
  \bibinfo{author}{\bibfnamefont{C.}~\bibnamefont{Rockstuhl}},
  \bibnamefont{and} \bibinfo{author}{\bibfnamefont{F.}~\bibnamefont{Lederer}},
  \bibinfo{journal}{Metamaterials} \textbf{\bibinfo{volume}{5}},
  \bibinfo{pages}{64} (\bibinfo{year}{2011}).

\bibitem[{\citenamefont{Mie}(1908)}]{mie1908beitrage}
\bibinfo{author}{\bibfnamefont{G.}~\bibnamefont{Mie}},
  \bibinfo{journal}{Annalen der Physik} \textbf{\bibinfo{volume}{330}},
  \bibinfo{pages}{377} (\bibinfo{year}{1908}).

\bibitem[{\citenamefont{Tribelsky and
  Lukyanchuk}(2006)}]{tribelsky2006anomalous}
\bibinfo{author}{\bibfnamefont{M.~I.} \bibnamefont{Tribelsky}}
  \bibnamefont{and} \bibinfo{author}{\bibfnamefont{B.~S.}
  \bibnamefont{Lukyanchuk}}, \bibinfo{journal}{Phys. Rev. Lett.}
  \textbf{\bibinfo{volume}{97}}, \bibinfo{pages}{263902}
  (\bibinfo{year}{2006}).

\bibitem[{\citenamefont{Barton et~al.}(1989)\citenamefont{Barton, Alexander,
  and Schaub}}]{barton1989theoretical}
\bibinfo{author}{\bibfnamefont{J.}~\bibnamefont{Barton}},
  \bibinfo{author}{\bibfnamefont{D.}~\bibnamefont{Alexander}},
  \bibnamefont{and} \bibinfo{author}{\bibfnamefont{S.}~\bibnamefont{Schaub}},
  \bibinfo{journal}{J. App. Phys.} \textbf{\bibinfo{volume}{66}},
  \bibinfo{pages}{4594} (\bibinfo{year}{1989}).

\bibitem[{\citenamefont{Almaas and Brevik}(1995)}]{almaas1995radiation}
\bibinfo{author}{\bibfnamefont{E.}~\bibnamefont{Almaas}} \bibnamefont{and}
  \bibinfo{author}{\bibfnamefont{I.}~\bibnamefont{Brevik}},
  \bibinfo{journal}{J. Opt. Soc. Am. B} \textbf{\bibinfo{volume}{12}},
  \bibinfo{pages}{2429} (\bibinfo{year}{1995}).

\bibitem[{\citenamefont{Fernandez-Corbaton
  et~al.}(2013)\citenamefont{Fernandez-Corbaton, Zambrana-Puyalto, Tischler,
  Vidal, Juan, and Molina-Terriza}}]{Fernandez-Corbaton2013c}
\bibinfo{author}{\bibfnamefont{I.}~\bibnamefont{Fernandez-Corbaton}},
  \bibinfo{author}{\bibfnamefont{X.}~\bibnamefont{Zambrana-Puyalto}},
  \bibinfo{author}{\bibfnamefont{N.}~\bibnamefont{Tischler}},
  \bibinfo{author}{\bibfnamefont{X.}~\bibnamefont{Vidal}},
  \bibinfo{author}{\bibfnamefont{M.~L.} \bibnamefont{Juan}}, \bibnamefont{and}
  \bibinfo{author}{\bibfnamefont{G.}~\bibnamefont{Molina-Terriza}},
  \bibinfo{journal}{Phys. Rev. Lett.} \textbf{\bibinfo{volume}{111}},
  \bibinfo{pages}{060401} (\bibinfo{year}{2013}).

\bibitem[{\citenamefont{Hulst and van~de Hulst}(1957)}]{hulst1957light}
\bibinfo{author}{\bibfnamefont{H.~C.} \bibnamefont{Hulst}} \bibnamefont{and}
  \bibinfo{author}{\bibfnamefont{H.~C.} \bibnamefont{van~de Hulst}},
  \emph{\bibinfo{title}{Light Scattering by Small Particles}}
  (\bibinfo{publisher}{Courier Corporation}, \bibinfo{year}{1957}).

\bibitem[{\citenamefont{Lee and El-Sharkawi}(2008)}]{lee2008modern}
\bibinfo{author}{\bibfnamefont{K.~Y.} \bibnamefont{Lee}} \bibnamefont{and}
  \bibinfo{author}{\bibfnamefont{M.~A.} \bibnamefont{El-Sharkawi}},
  \emph{\bibinfo{title}{Modern Heuristic Optimization Techniques}},
  vol.~\bibinfo{volume}{39} (\bibinfo{publisher}{John Wiley \& Sons},
  \bibinfo{year}{2008}).

\end{thebibliography}

\end{document}


\begin{center}
\textbf{\large{}Supplemental Material for:}
\par\end{center}{\large \par}

\begin{center}
\textit{\large{}``Fundamental Limits of Optical Force and Torque''}
\par\end{center}{\large \par}

\author{A. Rahimzadegan{$^{*1}$}, R. Alaee{$^{*1,2}$}, I. Fernandez-Corbaton{$^{1,3}$}, 
and C. Rockstuhl{$^{1,3}$}}

\affiliation{{\footnotesize{}{$^{1}$}Institute of Theoretical Solid State Physics,
Karlsruhe Institute of Technology, Karlsruhe, Germany}\\
{\footnotesize{} {$^{2}$}Max Planck Institute for the Science of
Light, Erlangen, Germany}\\
{\footnotesize{} {$^{3}$}Institute of Nanotechnology, Karlsruhe
Institute of Technology, Karlsruhe, Germany}\\
{\footnotesize{} $^{*}$Equally contributed to the work. Corresponding
authors: aso.rahimzadegan@kit.edu, rasoul.alaee@mpl.mpg.de }}

\maketitle
\tableofcontents{}

\setlength{\abovedisplayskip}{6pt} 
\setlength{\belowdisplayskip}{6pt}
\setlength{\abovedisplayshortskip}{6pt}
\setlength{\belowdisplayshortskip}{6pt}

\section{Multipole Expansion}

A macroscopically homogeneous scatterer is illuminated by an arbitrary
incident electromagnetic wave $\mathbf{E}_{\mathrm{i\mathrm{nc}}}$
and scatters the field $\mathbf{E}_{\mathrm{sca}}$. Based on the
multipole theory of electromagnetics, the scattered and incident fields
can be expanded in vector spherical wave functions (VSWF) as \cite{muhlig2011multipole,Xu1995c}:
\begin{eqnarray}
\mathbf{E}_{\mathrm{sca}}\left(\mathbf{r};\omega\right) & = & \sum_{j=1}^{\infty}\sum_{m=-j}^{j}E_{jm}\left[a_{jm}\left(\omega\right)\mathbf{N}_{jm}^{\left(3\right)}\left(\mathbf{r};\omega\right)+\mathrm{i}b_{jm}\left(\omega\right)\mathbf{M}_{jm}^{\left(3\right)}\left(\mathbf{r};\omega\right)\right],\label{eq:VSH-Expansion}\\
\mathbf{E}_{\mathrm{i\mathrm{nc}}}\left(\mathbf{r};\omega\right) & = & -\sum_{j=1}^{\infty}\sum_{m=-j}^{j}E_{jm}\left[p_{jm}\left(\omega\right)\mathbf{N}_{jm}^{\left(1\right)}\left(\mathbf{r};\omega\right)+\mathrm{i}q_{jm}\left(\omega\right)\mathbf{M}_{jm}^{\left(1\right)}\left(\mathbf{r};\omega\right)\right],\nonumber 
\end{eqnarray}
\noindent where $\mathbf{M}^{\left(f\right)}$ and $\mathbf{N}^{\left(f\right)}$
are the vector spherical wave functions (VSWFs) defined as:
\begin{eqnarray}
\mathbf{M}_{jm}^{\left(f\right)}\left(\mathbf{r};\omega\right) & = & \frac{z_{j}^{\left(f\right)}\left(kr\right)\exp\left(im\phi\right)}{\sqrt{j\left(j+1\right)}}\left[\mathrm{i}\pi_{jm}\left(\cos\theta\right)\mathbf{e}_{\theta}-\tau_{jm}\left(\cos\theta\right)\mathbf{e}_{\phi}\right],\\
\mathbf{N}_{jm}^{\left(f\right)}\left(\mathbf{r};\omega\right) & = & \sqrt{j\left(j+1\right)}\frac{z_{j}^{\left(f\right)}\left(kr\right)}{kr}P_{jm}\left(\cos\theta\right)\exp\left(im\phi\right)\mathbf{e}_{r}\\
 & + & \frac{1}{kr}\frac{d}{dr}\left[r\, z_{j}^{\left(f\right)}\left(kr\right)\right]\frac{\exp\left(im\phi\right)}{\sqrt{j\left(j+1\right)}}\left[\tau_{jm}\left(\cos\theta\right)\mathbf{e}_{\theta}+\mathrm{i}\pi_{jm}\left(\cos\theta\right)\mathbf{e}_{\phi}\right],\nonumber 
\end{eqnarray}
\noindent where $\pi_{jm}\left(\cos\theta\right)=m P_{jm}\left(\cos\theta\right)/\sin\theta,\,\,\tau_{jm}\left(\cos\theta\right)=d P_{jm}\left(\cos\theta\right)/d\theta$.
$z_{j}^{\left(1\right)}\left(kr\right)=j_{j}\left(kr\right)$ and $z_{j}^{\left(3\right)}\left(kr\right)=h_{j}^{\left(1\right)}\left(kr\right)$,
are the\textit{ spherical Bessel and Hankel functions of first kind},
respectively. The choice of $f$ should respect the physical
meaningfulness of the waves at the origin $\left(kr\rightarrow0\right)$
and infinity $\left(kr\rightarrow\infty\right)$. $P_{jm}$ are the
\textit{associated Legendre polynomials} defined as:
\begin{eqnarray}
P_{jm}\left(x\right) & = & \sqrt{\frac{\left(j+m\right)!}{\left(j-m\right)!}}\sqrt{\frac{\left(j-\left|m\right|\right)!}{\left(j+\left|m\right|\right)!}}\left[\mathrm{sign}\left(m\right)\right]^{m}\left(-1\right)^{\left|m\right|}\left(1-x^{2}\right)^{\frac{\left|m\right|}{2}}\frac{\mathrm{d}^{\left|m\right|}P_{j}\left(x\right)}{\mathrm{d}x^{\left|m\right|}},
\end{eqnarray}
\noindent where $P_{j}$ are the \textit{unassociated Legendre polynomials}.
$a_{jm}$ and $b_{jm}$ are the amplitudes of the VSWFs expanding the
scattered field. $p_{jm}$ and $q_{jm}$ are the amplitudes of the
VSWFs expanding the incident field. The subscripts $j$ and $m$ are
the total and z component of the total angular momentum number of
the interacting field. $k$ is the wavevector, and:
\begin{eqnarray*}
E_{jm} & = & \left|E_{0}\right|\sqrt{\frac{2j+1}{4\pi}}\sqrt{\frac{\left(j-m\right)!}{\left(j+m\right)!}}\left(-1\right)^{m},
\end{eqnarray*}
\noindent where $\left|E_{0}\right|$ is the magnitude of the expanded
wave.

\subsection*{The amplitudes of the VSWFs expanding the incident field}

If the incident field is known, then, their VSWF amplitudes can be
calculated, using orthogonality relations, by integrating the field
across a closed spherical surface of arbitrary radius $r$ as:
\begin{eqnarray}
p_{jm}\left(\omega\right) & = & -\frac{\intop_{0}^{2\pi}\mathrm{d}\phi\intop_{0}^{\pi}\mathbf{E}_{\mathrm{inc}}\left(r,\theta,\phi;\omega\right)\cdot\left[\mathbf{N}_{jm}^{\left(1\right)}\left(r,\theta,\phi;\omega\right)\right]^{*}\sin\theta\mathrm{d}\theta}{E_{jm}\intop_{0}^{2\pi}\mathrm{d}\phi\intop_{0}^{\pi}\left|\mathbf{N}_{jm}^{\left(1\right)}\left(r,\theta,\phi;\omega\right)\right|^{2}\sin\theta\mathrm{d}\theta},\label{eq:VSH-INC}\\
q_{jm}\left(\omega\right) & = & -\frac{\intop_{0}^{2\pi}\mathrm{d}\phi\intop_{0}^{\pi}\mathbf{\mathbf{E}_{\mathrm{inc}}}\left(r,\theta,\phi;\omega\right)\cdot\left[\mathbf{M}_{jm}^{\left(1\right)}\left(r,\theta,\phi;\omega\right)\right]^{*}\sin\theta\mathrm{d}\theta}{\mathrm{i}E_{jm}\intop_{0}^{2\pi}\mathrm{d}\phi\intop_{0}^{\pi}\left|\mathbf{M}_{jm}^{\left(1\right)}\left(r,\theta,\phi;\omega\right)\right|^{2}\sin\theta\mathrm{d}\theta}.
\end{eqnarray}
For an incident linearly x- and y-polarized plane wave propagating
in the z direction, based on the derivation of \cite{Xu1995c}, and
renormalization to the convention we have used in this letter, the
field VSWF amplitudes are as follows: 
\begin{eqnarray}
p_{jm}^{\mathrm{pw,x}}\left(\omega\right) & = & \mathrm{i}mq_{jm}^{\mathrm{pw,x}}\left(\omega\right)=-\mathrm{i}^{j+1}m\delta_{|m|1}\sqrt{\pi\left(2j+1\right)},\label{eq:Xpol}\\
p_{jm}^{\mathrm{pw,y}}\left(\omega\right) & = & \mathrm{i}mq_{jm}^{\mathrm{pw,y}}\left(\omega\right)=-\mathrm{i}^{j}\delta_{|m|1}\sqrt{\pi\left(2j+1\right)}\label{eq:Ypol}, 
\end{eqnarray}
\noindent where $\delta$ is the delta Kronecker. For a circularly
polarized plane wave propagating in the z direction $\mathbf{E}_{\mathrm{i\mathrm{nc}}}\left(\mathbf{r};\omega\right)=E_{\mathrm{0}}e^{\mathrm{i}kz}\left(\mathbf{e}_{x}+\mathrm{i}\sigma\mathbf{e}_{y}\right)/\sqrt{2},$
the incident field VSWF amplitudes are as follows ($\sigma$ is the
handedness of the wave): 
\begin{eqnarray}
p_{jm}^{\mathrm{cp}}\left(\omega\right) & = & \mathrm{i}mq_{jm}^{\mathrm{cp}}\left(\omega\right)=\frac{p_{jm}^{\mathrm{pw,x}}\left(\omega\right)+\mathrm{i}\sigma p_{jm}^{\mathrm{pw,y}}\left(\omega\right)}{\sqrt{2}}=-\mathrm{i}^{j+1}\sigma\delta_{m\sigma}\sqrt{2\pi\left(2j+1\right)}.\label{eq:Cpo}
\end{eqnarray}
For an arbitrarily polarized plane wave, propagating along z, the
VSWF amplitudes can be derived by a linear combination of the two linearly
x- and y-polarized polarized plane wave (Should be normalized).

\subsection*{The amplitudes of the VSWFs expanding the scattered field and Mie theory}

Finding the scattered field amplitudes are not a trivial task and
normally Maxwell equation solvers are required to calculate the scattered
field and derive the amplitudes through orthogonality relation, similar
to the case for the incident field VSWF amplitudes. However, for an
isotropic particle, analytical solutions exist, which is normally
refereed to as the Mie theory \cite{mie1908beitrage}. The equations
relating the scattered field VSWF amplitudes to that of the incident
field is:
\begin{eqnarray}
a_{jm}\left(\omega\right)=a_{j}\left(\omega\right)p_{jm}\left(\omega\right) & ~, ~& b_{jm}\left(\omega\right)=b_{j}\left(\omega\right)q_{jm}\left(\omega\right),\label{eq:Mie-c}
\end{eqnarray}
\noindent where $a_{j}$ and $b_{j}$ are known as the Mie coefficients.
The value of these coefficients for an isotropic sphere and core-multishell particle
can be found in \cite{Xu1995c,moroz2005recursive}. The spectrum of
the Mie coefficients consists of several resonance profiles, which
are the interaction channels for the light and the particle. Throughout
this letter, we refer to Mie coefficients as the Mie channels (MCs).
If the particle does not absorb light, at maximum light-particle interaction
(i.e. resonance), the Mie coefficients achieve the maximum value of
unity.

From now on, for simplicity, we ignore the angular argument $\omega$.

\subsection*{Spherical to Cartesian multipole conversion}

By comparison of the far field expressions of multipole moments and
the expanded fields of Eq.~\ref{eq:VSH-Expansion}, the induced electric
and magnetic Cartesian dipole moments can be related to the scattered
field VSWF amplitudes by \cite{muhlig2011multipole}:
\begin{eqnarray}
\mathbf{p} & = & \left(\begin{array}{c}
p_{x}\\
p_{y}\\
p_{z}
\end{array}\right)=C_{0}\left(\begin{array}{c}
a_{11}-a_{1-1}\\
\mathrm{i}\left(a_{11}+a_{1-1}\right)\\
-\sqrt{2}a_{10}
\end{array}\right),\label{eq:moments}\\
\mathbf{m} & = & \left(\begin{array}{c}
m_{x}\\
m_{y}\\
m_{z}
\end{array}\right)=cC_{0}\left(\begin{array}{c}
b_{11}-b_{1-1}\\
\mathrm{i}\left(b_{11}+b_{1-1}\right)\\
-\sqrt{2}b_{10}
\end{array}\right),
\end{eqnarray}
\noindent where $C_{0}=2\sqrt{3\pi}\mathrm{i}\left|E_{0}\right|k/cZ_{0}$,
$c$ is the speed of light, and $Z_{0}$ is the impedance of the free
space.

\section{Optical Cross Sections}

Scattering, extinction, and absorption cross sections are defined
as \cite{Bohren2008}:
\begin{eqnarray}
C_{\mathrm{sca}} & = & k^{-2}\sum_{n=1}^{\infty}\sum_{m=-j}^{j}\left(\left|a_{jm}\right|^{2}+\left|b_{jm}\right|^{2}\right),\\
C_{\mathrm{ext}} & = & k^{-2}\sum_{j=1}^{\infty}\sum_{m=-j}^{j}\Re\left(a_{jm}p_{jm}^{*}+b_{jm}q_{jm}^{*}\right),\\
C_{\mathrm{abs}} & = & C_{\mathrm{ext}}-C_{\mathrm{sca}}.\label{eq:Cabs}
\end{eqnarray}
Below the individual contribution of an electric MC ($a_{j}$) to
the optical cross sections of an \textit{isotropic} particle is calculated.

\subsection*{Scattering cross section}

The contribution of $a_{j}$ to the scattering cross section of the
particle when illuminated by an arbitrary illumination is: 
\begin{eqnarray}
C_{\mathrm{sca},j^{\mathrm{e}}} & = & k^{-2}\sum_{m=-n}^{n}\left|a_{jm}\right|^{2}=k^{-2}\left|a_{j}\right|^{2}\sum_{m=-j}^{j}\left|p_{jm}\right|^{2}.
\end{eqnarray}
If a z-propagating arbitrarily polarized plane wave (Eqs.~\ref{eq:Xpol}-\ref{eq:Cpo})
illumination is assumed, it simplifies to:
\begin{eqnarray}
C_{\mathrm{sca},j^{\mathrm{e}}} & = & \left(2j+1\right)\frac{\lambda^{2}}{2\pi}\left|a_{j}\right|^{2}.
\end{eqnarray}
To maximize the above expression, the particle should be non-absorptive
and the frequency of the light should be tuned for the resonance of
the MC, where $a_{j}=1$. Therefore, the maximum contribution is equal
to \cite{ruan2010superscattering}:
\begin{eqnarray}
\left[C_{\mathrm{sca},j^{\mathrm{e}}}\right]_{\mathrm{max}} & = & \left(2j+1\right)\frac{\lambda^{2}}{2\pi}.\label{CscaMAx}
\end{eqnarray}

\subsection*{Extinction cross section}

The contribution of $a_{j}$ to the extinction cross section of the
particle when illuminated by an arbitrary illumination is: 
\begin{eqnarray}
C_{\mathrm{ext},j^{e}} & = & k^{-2}\sum_{m=-j}^{j}\Re\left(a_{jm}p_{jm}^{*}\right)=k^{-2}\Re\left(a_{j}\right)\sum_{m=-j}^{j}\left|p_{jm}\right|^{2}.
\end{eqnarray}
If a z-propagating arbitrarily polarized plane wave (Eqs.~\ref{eq:Xpol}-\ref{eq:Cpo})
illumination is assumed, it simplifies to:
\begin{eqnarray}
C_{\mathrm{ext},j^{\mathrm{e}}} & = & \left(2j+1\right)\frac{\lambda^{2}}{2\pi}\Re\left(a_{j}\right).\label{eq:Cext}
\end{eqnarray}
Therefore, the maximum contribution occurs at the MC resonance of
a non-absorbing particle and is equal to:
\begin{eqnarray}
\left[C_{\mathrm{ext},j^{\mathrm{e}}}\right]_{\mathrm{max}} & = & \left(2j+1\right)\frac{\lambda^{2}}{2\pi}.\label{CextMax}
\end{eqnarray}
Extinction cross section characterizes the total interaction of the
particle with the light, i.e. the scattering plus the absorption cross
section. When the non-radiative loss is very small, then, at resonance
the extinction cross section maximizes. We refer to this regime as
the over-coupling regime. In the over-coupling region, the absorption
cross section vanishes and the scattering and extinction cross sections
are both equal.

\subsection*{Absorption cross section}

Finally, the contribution of $a_{j}$ to the absorption cross section
of the particle is: 
\begin{eqnarray}
C_{\mathrm{abs},j^{e}} & = & C_{\mathrm{ext,}j^{e}}-C_{\mathrm{sca},j^{e}}=\frac{\left[\Re\left(a_{j}\right)-\left|a_{j}\right|^{2}\right]}{k^{2}}\sum_{m=-j}^{j}\left|p_{jm}\right|^{2}=\left(2j+1\right)\frac{\lambda^{2}}{2\pi}\left[\Re\left(a_{j}\right)-\left|a_{j}\right|^{2}\right].\label{Cabs}
\end{eqnarray}
For a non-absorbing particle, the cross section vanishes and $\Re\left(a_{j}\right)=\left|a_{j}\right|^{2}$.
It can be shown that the maximum of $\Re\left(a_{j}-\left|a_{j}\right|^{2}\right)$
is at the channel resonance and is equal to 0.25. Therefore, the maximum
absorption cross section is \cite{miroshnichenko2016ultimate}:
\begin{eqnarray}
\left[C_{\mathrm{abs},j^{\mathrm{e}}}\right]_{\mathrm{max}} & = & \frac{\left(2j+1\right)}{4}\frac{\lambda^{2}}{2\pi}.\label{eq:CabsMax}
\end{eqnarray}
We refer to this regime as the critical coupling, where non-radiative
loss ($\gamma_{\mathrm{nr}}$) is equal to the radiative loss ($\gamma_{\mathrm{r}}$).
In critical coupling regime, the scattering and absorption cross section
are equal.

Similar results derive for the individual contribution of a magnetic
MC $b_{j}$ to the optical cross sections:
\begin{eqnarray}
C_{\mathrm{sca},j^{\mathrm{m}}} & = & \left(2j+1\right)\frac{\lambda^{2}}{2\pi}\left|b_{j}\right|^{2},\\
C_{\mathrm{ext},j^{\mathrm{m}}} & = & \left(2j+1\right)\frac{\lambda^{2}}{2\pi}\Re\left(b_{j}\right),\\
C_{\mathrm{abs},j^{\mathrm{m}}} & = & \left(2j+1\right)\frac{\lambda^{2}}{2\pi}\left[\Re\left(b_{j}\right)-\left|b_{j}\right|^{2}\right],\\
\left[C_{\mathrm{sca},j^{\mathrm{m}}}\right]_{\mathrm{max}} & = & \left(2j+1\right)\frac{\lambda^{2}}{2\pi},\\
\left[C_{\mathrm{ext},j^{\mathrm{m}}}\right]_{\mathrm{max}} & = & \left(2j+1\right)\frac{\lambda^{2}}{2\pi},\\
\left[C_{\mathrm{abs},j^{\mathrm{m}}}\right]_{\mathrm{max}} & = & \frac{\left(2j+1\right)}{4}\frac{\lambda^{2}}{2\pi}.
\end{eqnarray}
Note that the above-mentioned upper bounds that are vastly
used in the literature are only valid given the assumptions: isotropy
of the particle, individual contribution of a MC and plane wave illumination.
Breaking any of these assumptions might change the upper bounds. For
example a non-absorbing dipolar dual sphere can scatter twice as
the upper bound derived in Eq.~\ref{CscaMAx}. This is because the
magnetic and electric MCs overlap spectrally and the total
cross section is the sum of the two. The total optical cross section
is an incoherent sum of individual MC contributors and no interference
terms appear if multiple MCs profiles overlap spectrally.

\section{Optical Force}

\subsection*{Spherical multipole expansion}

The optical force exerted on an arbitrary particle by an arbitrary
illumination, expanded into spherical multipoles, is as follows (the
formula is what \cite{barton1989theoretical,almaas1995radiation}
have derived, but re-normalized to be consistent with other formulas
used in this letter) (SI units):
\textit{
\begin{eqnarray}
\mathbf{F} & = & F_{x}\mathbf{e}_{x}+F_{y}\mathbf{e}_{y}+F_{z}\mathbf{e}_{z}, \\
F_{z} & = & -\frac{I_{0}}{c}\frac{\lambda^{2}}{4\pi^{2}}\,\sum_{j=1}^{\infty}\sum_{m=-j}^{j}\Im\left[\frac{\left(j+2\right)}{\left(j+1\right)}\sqrt{\frac{\left(j+1\right)^{2}-m^{2}}{\left(2j+1\right)\left(2j+3\right)}}\left(2a_{j+1,m}a_{jm}^{*}\right.\right.\nonumber \\
 &  & \left.-a_{j+1,m}p_{jm}^{*}-p_{j+1,m}a_{jm}^{*}+2b_{j+1,m}b_{jm}^{*}-b_{j+1,m}q_{jm}^{*}-q_{j+1,m}b_{jm}^{*}\right)\nonumber \\
 &  & \left.+\frac{m}{j\left(j+1\right)}\left(2a_{jm}b_{jm}^{*}-a_{jm}q_{jm}^{*}-p_{jm}b_{jm}^{*}\right)\right],\label{eq:Barton}\\
F_{x}+\mathrm{i}F_{y} & = & -\frac{I_{0}}{c}\frac{\lambda^{2}}{4\pi^{2}}\sum_{j=1}^{\infty}\sum_{m=-j}^{j}\Biggl\{\sqrt{\frac{\left(j+m+2\right)\left(j+m+1\right)}{\left(2j+1\right)\left(2j+3\right)}}\frac{\left(j+2\right)}{\left(j+1\right)}\biggl[2a_{jm}a_{j+1,m+1}^{*}\nonumber \\
 &  & -a_{jm}p_{j+1,m+1}^{*}-p_{jm}a_{j+1,m+1}^{*}+2b_{jm}b_{j+1,m+1}^{*}-b_{jm}q_{j+1,m+1}^{*}\nonumber \\
 &  & -q_{jm}b_{j+1,m+1}^{*}\biggr]+\sqrt{\frac{\left(j-m+1\right)\left(j-m+2\right)}{\left(2j+1\right)\left(2j+3\right)}}\frac{\left(j+2\right)}{\left(j+1\right)}\biggl[2a_{j+1,m-1}a_{jm}^{*}\nonumber \\
 &  & -a_{j+1,m-1}p_{jm}^{*}-p_{j+1,m-1}a_{jm}^{*}+2b_{j+1,m-1}b_{jm}^{*}-b_{j+1,m-1}q_{jm}^{*}\nonumber \\
 &  & -q_{j+1,m-1}b_{jm}^{*}\biggr]-\frac{\sqrt{\left(j+m+1\right)\left(j-m\right)}}{j\left(j+1\right)}\biggl[-2a_{jm}b_{j,m+1}^{*}+2b_{jm}a_{j,m+1}^{*}\nonumber \\
 &  & +a_{jm}q_{j,m+1}^{*}-b_{jm}p_{j,m+1}^{*}-q_{jm}a_{j,m+1}^{*}+p_{jm}b_{j,m+1}\biggr]\Biggr\}.
\end{eqnarray}
}
For an isotropic particle, and z-propagating plane wave illumination,
lateral force is always zero, i.e., $F_{x}=F_{y}=0$. Therefore, only
the parallel force $F_{z}$ is relevant and throughout the letter,
it is referred to as the exerted optical force $F$. Assuming isotropy
of the particle and rearranging the formula with respect to the Mie
coefficients, the exerted optical force by an arbitrary illumination
can be written as:
\begin{eqnarray}
F & = & \sum_{j=1}^{\infty}\left[F_{j^{\mathrm{e}}}+F_{j^{\mathrm{m}}}+F_{j^{\mathrm{e}}j^{\mathrm{m}}}+F_{j^{\mathrm{e}}\left(j+1\right)^{\mathrm{e}}}+F_{j^{\mathrm{m}}\left(j+1\right)^{\mathrm{m}}}\right].\label{eq:force-n-1-2}
\end{eqnarray}

\subsection*{Fundamental bounds for the single constituents of the total optical force}

The optical force consists of various terms, showing different MC contributions. The individual
contribution of an electric $(F_{j^{e}})$/ magnetic $(F_{j^{m}})$
MC, the contribution of the interference of the electric and magnetic
MCs $(F_{j^{\mathrm{e}}j^{\mathrm{m}}})$, the contribution of the
interference of two electric MCs with adjacent $j$ numbers $\left[F_{j^{e}\left(j+1\right)^{e}}\right]$,
and the contribution of the interference of two magnetic MCs with
adjacent $j$ numbers $\left[F_{j^{m}\left(j+1\right)^{m}}\right]$.

The normalizing force factor, that will be used throughout the letter,
is defined as:
\begin{eqnarray}
F^{\mathrm{norm}} & = & \frac{I_{0}}{c}\frac{\lambda^{2}}{2\pi}.\label{eq:}
\end{eqnarray}
Below each of the terms are analytically calculated and the
upper bounds for their contribution are derived.

\paragraph*{Individual contribution of an electric $(F_{j^{\mathrm{e}}})$/magnetic
$(F_{j^{\mathrm{m}}})$ MC:}

Ignoring the interference terms, the individual contribution of an
induced electric multipole to the exerted optical force by an arbitrary
illumination is derived as: 
\textit{
\begin{eqnarray}
F_{j^{e}} & = & \frac{I_{0}}{c}\frac{\lambda^{2}}{4\pi^{2}}\,\sum_{m=-j}^{j}\Im\left[\frac{\left(j+2\right)}{\left(j+1\right)}\sqrt{\frac{\left(j+1\right)^{2}-m^{2}}{\left(2j+1\right)\left(2j+3\right)}}p_{j+1,m}a_{jm}^{*}\right.\nonumber\\
 &  & \left.+\frac{\left(j-1\right)}{j}\sqrt{\frac{\left(j-m\right)\left(j+m\right)}{\left(2j-1\right)\left(2j+1\right)}}a_{jm}p_{j-1,m}^{*}+\frac{m}{j\left(j+1\right)}a_{jm}q_{jm}^{*}\right]\nonumber \\
 & = & \frac{I_{0}}{c}\frac{\lambda^{2}}{4\pi^{2}}\,\sum_{m=1,-1}\Im\left[\frac{\left(j+2\right)}{\left(j+1\right)}\sqrt{\frac{\left(j+1\right)^{2}-m^{2}}{\left(2j+1\right)\left(2j+3\right)}}p_{j+1,m}\left(a_{j}^{*}p_{jm}^{*}\right)\right.\nonumber \\
 &  & \left.+\frac{\left(j-1\right)}{j}\sqrt{\frac{\left(j^{2}-m^{2}\right)}{\left(4j^{2}-1\right)}}a_{j}p_{jm}p_{j-1m}^{*}+\frac{m}{j\left(j+1\right)}a_{j}p_{jm}q_{jm}^{*}\right]\nonumber \\
 & = & \frac{I_{0}}{c}\frac{\lambda^{2}}{4\pi^{2}}\Im\left\{ a_{j}\sum_{m=-j}^{j}p_{jm}\left[-\frac{\left(j+2\right)\sqrt{\left(j+1\right)^{2}-m^{2}}}{\left(j+1\right)\sqrt{\left(2j+1\right)\left(2j+3\right)}}p_{j+1,m}^{*}\right.\right.\nonumber \\
 &  & \left.\left.+\frac{\left(j-1\right)}{j}\sqrt{\frac{\left(j^{2}-m^{2}\right)}{\left(4j^{2}-1\right)}}p_{j-1m}^{*}+\frac{m}{j\left(j+1\right)}q_{jm}^{*}\right]\right\} .
\end{eqnarray}
}
For a z-propagating arbitrarily polarized plane wave illumination,
using Eq.~\ref{eq:Xpol} and Eq.~\ref{eq:Cext}, it can be written
as:
\textit{
\begin{eqnarray}
F_{j^{\mathrm{e}}} & = & \frac{I_0}{c}C_{\mathrm{ext,}j^{e}}=\left(2j+1\right)F^{\mathrm{norm}}\Re\left(a_{j}\right).
\end{eqnarray}
}
Therefore, at resonance of the electric MC, for a non-absorbing particle,
the maximum force contribution is:
\begin{eqnarray}
\left[F_{j^{\mathrm{e}}}\right]_{\mathrm{max}} & = & \left(2j+1\right)\frac{I_{0}}{c}\frac{\lambda^{2}}{2\pi}=\left(2j+1\right)F^{\mathrm{norm}}.
\end{eqnarray}
Similarity, it can be proven that the maximum individual contribution
of a magnetic MC is:
\begin{eqnarray}
\left[F_{j^{\mathrm{m}}}\right]_{\mathrm{max}} & = & \frac{I_0}{c}\left[C_{\mathrm{ext,}j^{m}}\right]_{\mathrm{max}}=\left(2j+1\right)\frac{I_{0}}{c}\frac{\lambda^{2}}{2\pi}=\left(2j+1\right)F^{\mathrm{norm}}.
\end{eqnarray}

\paragraph*{Contribution due to the interference of electric and magnetic MC
with the same j $(F_{j^{e}j^{m}})$:}

The partial contribution of the interference of $a_{j}$ and $b_{j}$
is:
\textit{
\begin{eqnarray}
F_{j^{\mathrm{e}}j^{\mathrm{m}}} & = & -\frac{2\lambda^{2} I_0 }{4\pi^{2}cj\left(j+1\right)}\sum_{m=-j}^{j}m\,\Im\left(a_{jm}b_{jm}^{*}\right)=-\frac{I_{0} \lambda^{2}}{2\pi^{2} cj\left(j+1\right)}\sum_{m=-j}^{j}m\,\Im\left(a_{j}b_{j}^{*}p_{jm}q_{jm}^{*}\right).
\end{eqnarray}
}
For a z-propagating arbitrarily polarized plane wave illumination,
it simplifies to:
\begin{eqnarray}
F_{j^{\mathrm{e}}j^{\mathrm{m}}} & = & -F^{\mathrm{norm}}\frac{\left(2j+1\right)}{j\left(j+1\right)}\sum_{m=-j}^{j}m^{2}\Im\left(a_{j}b_{j}^{*}\mathrm{i}\delta_{|m|1}\right)=-\frac{2\left(2j+1\right)}{j\left(j+1\right)}\Re\left(a_{j}b_{j}^{*}\right)F^{\mathrm{norm}}.
\end{eqnarray}
The maximum contribution is when $a_{j}$ and $b_{j}$ simultaneously
resonate and the particle is non-absorptive:
\textit{
\begin{eqnarray}
\left[F_{j^{\mathrm{e}}j^{\mathrm{m}}}\right]_{\mathrm{max}} & = & -\frac{2\left(2j+1\right)}{j\left(j+1\right)}F^{\mathrm{norm}}.
\end{eqnarray}
}

\paragraph*{Contribution of the interference of two electric $\left[F_{j^{\mathrm{e}}\left(j+1\right)^{\mathrm{e}}}\right]$/
two magnetic $\left[F_{j^{\mathrm{m}}\left(j+1\right)^{\mathrm{m}}}\right]$
MCs with adjacent $j$:}

The partial contribution of the interference of $a_{j}$ and $a_{j+1}$
is:
\textit{
\begin{eqnarray}
F_{j^{e}\left(j+1\right)^{e}} & = & -\frac{I_{0}}{c}\frac{\lambda^{2}}{4\pi^{2}}\,\sum_{m=-j}^{j}\Im\left[\frac{\left(j+2\right)}{\left(j+1\right)}\sqrt{\frac{\left(j+1\right)^{2}-m^{2}}{\left(2j+1\right)\left(2j+3\right)}}2a_{j+1,m}a_{jm}^{*}\right]\nonumber \\
 & = & -\frac{I_{0}}{c}\frac{\lambda^{2}}{2\pi^{2}}\frac{\left(j+2\right)}{\left(j+1\right)}\sqrt{\frac{j\left(j+2\right)}{\left(2j+1\right)\left(2j+3\right)}}\sum_{m=-j}^{j}\Im\left[p_{j+1,m}p_{jm}^{*}a_{j+1}a_{j}^{*}\right].
\end{eqnarray}
}
For a z-propagating arbitrarily polarized plane wave illumination,
it simplifies to:
\begin{eqnarray}
F_{j^{\mathrm{e}}\left(j+1\right)^{\mathrm{e}}} & = & -\frac{2j\left(j+2\right)}{\left(j+1\right)}\Re\left(a_{j+1}a_{j}^{*}\right)F^{\mathrm{norm}}.
\end{eqnarray}
The maximum contribution is when an electric MC resonance spectrally
overlaps with an electric MC resonance of one order higher and the
particle is non-absorptive:
\textit{
\begin{eqnarray}
\left[F_{j^{\mathrm{e}}\left(j+1\right)^{\mathrm{e}}}\right]_{\mathrm{max}} & = & -\frac{2j\left(j+2\right)}{\left(j+1\right)}F^{\mathrm{rorm}}.
\end{eqnarray}
}
Similarly:
\textit{
\begin{eqnarray}
F_{j^{\mathrm{m}}\left(j+1\right)^{\mathrm{m}}} & = & -\frac{2j\left(j+2\right)}{\left(j+1\right)}\Re\left(b_{j+1}b_{j}^{*}\right)F^{\mathrm{norm}},
\end{eqnarray}
}
\noindent and the maximum value can be derived for simultaneous resonance
of $b_{j}$ and $b_{j+1}$ for a non-absorbing particle:
\textit{
\begin{eqnarray}
\left[F_{j^{\mathrm{m}}\left(j+1\right)^{\mathrm{m}}}\right]_{\mathrm{max}} & = & -\frac{2j\left(j+2\right)}{\left(j+1\right)}F^{\mathrm{norm}}.
\end{eqnarray}
}

\paragraph*{Maximum positive Contribution of the interference terms:}
For the non-interference terms, the maximum contribution is always positive and the terms cannot be negative in the whole spectrum. On the other hand, the maximum contribution of the interference terms are always negative. However, as shown in Fig.~2, the interference terms can also be positive at some frequencies. Here, we want to find the maximum positive contribution for the interference terms. All interference terms in the simplified form (isotropic particle, plane wave illumination), can be written as:
\begin{eqnarray}
F_{j^{\mathrm{e}}j^{\mathrm{m}}} & = & -\frac{2\left(2j+1\right)}{j\left(j+1\right)}\Re\left(a_{j}b_{j}^{*}\right)F^{\mathrm{norm}},\\
F_{j^{\mathrm{e}}\left(j+1\right)^{\mathrm{e}}} & = & -\frac{2j\left(j+2\right)}{\left(j+1\right)}\Re\left(a_{j+1}a_{j}^{*}\right)F^{\mathrm{norm}},\\
F_{j^{\mathrm{m}}\left(j+1\right)^{\mathrm{m}}} & = & -\frac{2j\left(j+2\right)}{\left(j+1\right)}\Re\left(b_{j+1}b_{j}^{*}\right)F^{\mathrm{norm}}.
\end{eqnarray}
Since in general, the MCs are independent, it will be enough to calculate the extema of $\Re\left(u v^{*}\right)$, where $u$ and $v$ are two arbitrary independent MCs. For the Mie coeficients of a non-absorbing particle, as shown before $\Re\left(u\right)=\left|u\right|^{2}$.
For any complex number $u  =  |u|e^{\mathrm{i}\alpha}$. Therefore:
\begin{eqnarray}
\Re\left(|u|e^{\mathrm{i}\alpha}\right) & = & |u|^{2}\nonumber\\
|u|\cos\alpha & = & |u|^{2}\nonumber\\
\cos\alpha & = & |u|,
\end{eqnarray}

\noindent where $-\frac{\pi}{2}\leq\alpha\leq\frac{\pi}{2}$. As a result, for any non-absorbing particle, the Mie coefficients can be written in the form \cite{hulst1957light}:
\hspace{0.1cm}
\begin{eqnarray}
u & = & \cos\alpha e^{\mathrm{i}\alpha}.\label{eq:MieCoff}
\end{eqnarray}
\hspace{0.1cm}

We want to find the maximum and minimum of the function $f=\Re\left(u v^{*}\right)$,
which appears in the interference contributions to the optical force. It can be written as:	
\begin{eqnarray}
f & = & \Re\left(u v^{*}\right)\nonumber\\
 & = & \Re\left(\cos\alpha e^{\mathrm{i}\alpha}\cos\beta e^{-\mathrm{i}\beta}\right)\nonumber\\
 & = & \cos\alpha\cos\beta\cos\left(\alpha-\beta\right).
\end{eqnarray}
To find the extrema of the function, derivative of the function with respect to $\alpha$ and $\beta$ is required. The maximum of the function is trivial and is equal to 1 (at the simultaneous resonance of the two MCs), the minimum of the function can be derived as -1/8 (off-resonance of the two MCs):
\begin{eqnarray}
f (\alpha=\pm\frac{\pi}{3},\beta=\mp\frac{\pi}{3}) & = & -\frac{1}{8}.
 \end{eqnarray}
In summary, the fundamental bounds for the different constituents are:
\begin{eqnarray}
0\leq & F_{j^{\mathrm{e}}} & \leq\left(2j+1\right)F^{\mathrm{norm}},\\
0\leq & F_{j^{\mathrm{m}}} & \leq\left(2j+1\right)F^{\mathrm{norm}},\\
-\frac{2\left(2j+1\right)}{j\left(j+1\right)}F^{\mathrm{norm}}\leq & F_{j^{\mathrm{e}}j^{\mathrm{m}}} & \leq\frac{\left(2j+1\right)}{4j\left(j+1\right)}F^{\mathrm{norm}},\\
-\frac{2j\left(j+2\right)}{\left(j+1\right)}F^{\mathrm{norm}}\leq & F_{j^{\mathrm{e}}\left(j+1\right)^{\mathrm{e}}} & \leq\frac{j\left(j+2\right)}{4\left(j+1\right)}F^{\mathrm{norm}},\\
-\frac{2j\left(j+2\right)}{\left(j+1\right)}F^{\mathrm{norm}}\leq & F_{j^{\mathrm{m}}\left(j+1\right)^{\mathrm{m}}} & \leq\frac{j\left(j+2\right)}{4\left(j+1\right)}F^{\mathrm{norm}}.
\end{eqnarray}

\subsection*{Fundamental bounds for the total optical force}

Now that for all the contributions to the total optical force, we have found the fundamental bounds, it is time to calculate the bounds for the total optical force itself. For that goal, we will write all Mie coefficients as the form introduced in Eq.~\ref{eq:MieCoff}. For the Mie coefficients $a_j$ and $b_j$, we write:
\hspace{0.1cm}
\begin{eqnarray}
a_j & = & \cos\alpha_j e^{\mathrm{i}\alpha_j},\label{eq:MieCoffa}\\
b_j & = & \cos\beta_j e^{\mathrm{i}\beta_j}.\label{eq:MieCoffa}
\end{eqnarray}
\hspace{0.1cm}
Therefore the total optical force in Eq.~\ref{eq:force-n-1-2} can be written as:
\begin{eqnarray}
F & = & F^{\mathrm{norm}} \sum_{j=1}^{\infty}\left[(2j+1)\Re(a_j+b_j)-\frac{2(2j+1)}{j(j+1)}\Re(a_j b_j^*)-\frac{2j(j+2)}{j+1}\Re(a_j a_{j+1}^* +b_j b_{j+1}^*)\right]\nonumber\\
 & = & F^{\mathrm{norm}}\{ \sum_{j=1}^{\infty} (2j+1)(\cos^2\alpha_j+\cos^2\beta_j)-\frac{2(2j+1)}{j(j+1)}\cos \alpha_j \cos \beta_j \cos(\alpha_j-\beta_j)\nonumber\\
 &  & -\frac{2j(j+2)}{j+1}[\cos \alpha_j \cos \alpha_{j+1} \cos(\alpha_j-\alpha_{j+1})+\cos \beta_j \cos \beta_{j+1} \cos(\beta_j-\beta_{j+1})]\}.
\end{eqnarray}

At a specific frequency, the angles $\alpha_j$ and $\beta_j$ can be assume to be independent of each other. This independence is assumed based on the degree of freedom that the design of the particle gives 
(e.g the permittivity and radii of a core-multishell particle). Here, assuming that the particle is excitable up to the multiple response $j_\mathrm{max}$, the total optical force is rewritten as the following form:
\begin{eqnarray}
F & = & F^{\mathrm{norm}} \{\sum_{j=1}^{j_\mathrm{max}} (2j+1)(\cos^2\alpha_j+\cos^2\beta_j)-\frac{2(2j+1)}{j(j+1)}\cos \alpha_j \cos \beta_j \cos(\alpha_j-\beta_j)\nonumber\\
 &  &  -\sum_{j=1}^{j_\mathrm{max}-1} \frac{2j(j+2)}{j+1}[\cos \alpha_j \cos \alpha_{j+1} \cos(\alpha_j-\alpha_{j+1})+\cos \beta_j \cos \beta_{j+1} \cos(\beta_j-\beta_{j+1})]\}.
\end{eqnarray}

For the simplest case of a dipolar approximation (i.e. only the electric and magnetic dipole moments), the formula developed above for the total optical force is simplified into:
\begin{eqnarray}
F_\mathrm{dipole} & = & 3F^{\mathrm{norm}} [ \cos^2\alpha_1+\cos^2\beta_1-\cos \alpha_1 \cos \beta_1 \cos(\alpha_1-\beta_1)].
\end{eqnarray}
Previously the maximum individual contribution of either of the electric or magnetic dipoles are calculated to be $3F^{\mathrm{norm}}$. The total optical force for a dual particle at resonance is also  $3F^{\mathrm{norm}}$ (Fig.~\ref{fig:Non-absorptive_dual_dielectric}). Can the force on a dipolar particle, surpass those values? The function above can be optimized numerically and analytically. The maximum of the function is calculated to be:
\begin{eqnarray}
(F_\mathrm{dipole})_\mathrm{max}|_{\alpha_1=\pm\frac{\pi}{6},\beta_1=\mp\frac{\pi}{6}} & = & \frac{27}{8}F^{\mathrm{norm}}.
\end{eqnarray}
In Fig.~3(c), using particle swarm optimization (PSO), we have optimized a core-multishell particle to result in the maximum possible normalized total optical force in dipolar approximation. Details are given in a separate section below. The maximum of the function can be analytically derived up to $j_\mathrm{max}=3$. 
 To solve for higher orders $j_\mathrm{max}>3$, a genetic algorithm is implemented to maximize the function $F/F^\mathrm{norm}$ with $2j_\mathrm{max}$ variables ($\alpha_j,\beta_j~~j=1,2,..,j_\mathrm{max}$). The maximum normalized total optical force as a function of  $j_\mathrm{max}$ up to $j_\mathrm{max}=10$ is 
shown in Fig.~3(a). The maximum values from $j_\mathrm{max}=1$ to $j_\mathrm{max}=10$ are calculated to be:
\begin{eqnarray}
(F_\mathrm{Total})_\mathrm{max} & = & \{3.37,9.18,17.11,27.07,39.05,53.03,69.03,87.02,105.41,128.18,... \} F^\mathrm{norm}.
\end{eqnarray}
The maximum of the total optical force as a function of $j_\mathrm{max}$ is plotted up to $j_\mathrm{max}=50$. A basic fitting to a quadratic function results to the following equation:
\begin{eqnarray}
(F_\mathrm{Total})_\mathrm{max} & = & (0.97631~ j^2_\mathrm{max}+3.1173 ~j_\mathrm{max}-0.888333) F^\mathrm{norm},
\end{eqnarray}
\noindent where the norm of residuals is equal to 1.0699.

\subsection*{Cartesian dipole expansion}

The time averaged optical force exerted on a dipolar particle by an
arbitrary incident wave, expanded into Cartesian dipole moments, is
\cite{Nieto-Vesperinas2010,chen2011optical} (SI units) :
\begin{eqnarray}
\mathbf{F} & = & \mathbf{F}_{p}+\mathbf{F}_{m}+\mathbf{F}_{pm},\label{eq:Force_moments}\\
\mathbf{F}_{p} & = & \frac{1}{2}\Re\left[\left(\nabla\mathbf{E}_{\mathrm{inc}}^{\mathrm{*}}\right)\cdot\mathbf{p}\right],\\
\mathbf{F}_{m} & = & \frac{1}{2}\Re\left[\left(\nabla\mathbf{B}_{\mathrm{inc}}^{\mathrm{*}}\right)\cdot\mathbf{m}\right],\\
\mathbf{F}_{pm} & = & -\frac{Z_{0}k^{4}}{12\pi}\Re\left[\left(\mathbf{p}\times\mathbf{m}^{*}\right)\right],
\end{eqnarray}
\noindent where $\mathbf{F}_{\mathrm{p}}$, $\mathbf{F}_{\mathrm{m}}$
and $\mathbf{F}_{\mathrm{pm}}$ are the forces due to the induced
electric Cartesian dipole moment $\mathrm{\mathbf{p}}$, magnetic
Cartesian dipole moment $\mathrm{\mathbf{m}}$, and the interference
of the two, respectively. The gradient of a vector $\mathbf{A}$ is
defined as:
\begin{eqnarray}
\nabla\mathbf{A} & = & \left[\begin{array}{ccc}
\partial_{\mathit{x}}A_{\mathit{x}} & \partial_{\mathit{x}}A_{\mathit{y}} & \partial_{\mathit{x}}A_{\mathit{z}}\\
\partial_{\mathit{y}}A_{\mathit{x}} & \partial_{\mathit{y}}A_{\mathit{y}} & \partial_{\mathit{y}}A_{\mathit{z}}\\
\partial_{\mathit{z}}A_{\mathit{x}} & \partial_{\mathit{z}}A_{\mathit{y}} & \partial_{\mathit{z}}A_{\mathit{z}}
\end{array}\right].
\end{eqnarray}

\section{Optical Torque}

\subsection*{Spherical multipole expansion}

The optical torque exerted on an arbitrary scatterer in free space,
using the VSWF amplitudes, is as follows (re-normalized) (notice that
there is a missing $j$ in the reference paper \cite{barton1989theoretical},
in the original form) (SI units):
\begin{eqnarray}
\mathbf{N} & = & N_{x}\mathbf{e}_{x}+N_{y}\mathbf{e}_{y}+N_{z}\mathbf{e}_{z},\\
N_{z} & = & -\frac{I_{0}}{c}\frac{\lambda^{3}}{8\pi^{3}}\sum_{j=1}^{\infty}\sum_{m=-j}^{j}m\left[\left|a_{jm}\right|^{2}+\left|b_{jm}\right|^{2}-\Re\left(a_{jm}p_{jm}^{*}+b_{jm}q_{jm}^{*}\right)\right],\\
N_{x}+\mathrm{i}N_{y} & = & -\frac{I_{0}}{c}\frac{\lambda^{3}}{8\pi^{3}}\sum_{j=1}^{\infty}\sum_{m=-j}^{j}\Biggl\{\sqrt{\left(j+m+1\right)\left(j-m\right)}\biggl[a_{jm}a_{j,m+1}^{*}\nonumber\\
 &  & +b_{jm}b_{j,m+1}^{*}-\frac{1}{2}\left(a_{jm}p_{j,m+1}^{*}+a_{j,m+1}^{*}p_{jm}+b_{j,m}q_{j,m+1}^{*}+q_{jm}b_{j,m+1}^{*}\right)\biggr]\Biggl\}.
\end{eqnarray}
For a z propagating plane wave, the only non-zero torque is $N_{z}$
and it will be refereed to as the exerted optical torque $N$.

The exerted optical torque on an isotropic particle by an arbitrary
illumination can be rewritten as:
\noindent 
\begin{eqnarray}
N & = & \sum_{j=1}^{\infty}\left(N_{j^{e}}+N_{j^{m}}\right).\label{eq:force-n-1-1}
\end{eqnarray}
As can be seen, for the optical torque along $\mathbf{k}$ direction (i.e. the z-direction),
no interference term contributes.

We define the normalizing torque factor as:
\begin{eqnarray}
N^{\mathrm{norm}} & = & \frac{I_{0}}{\omega}\frac{\lambda^{2}}{8\pi}.\label{eq:-1}
\end{eqnarray}
Below each of the two terms are analytically calculated and
the upper bounds for their contribution are derived.

\subsection*{Fundamental bounds for the single constituents of the total optical torque}

\paragraph*{Contribution of an individual electric $(N_{j^{e}})$/magnetic $(N_{j^{m}})$
MC:}

The individual contribution of an electric MC $a_{j}$ is:
\begin{eqnarray}
N_{j^{e}} & = & -\frac{I_{0}}{c}\frac{\lambda^{3}}{8\pi^{3}}\sum_{m=-j}^{j}m\left[\left|a_{jm}\right|^{2}-\Re\left(a_{jm}p_{jm}^{*}\right)\right]\nonumber \\
 & = & -\frac{I_{0}}{c}\frac{\lambda^{3}}{8\pi^{3}}\sum_{m=-j}^{j}m\left[\left|a_{j}\right|^{2}\left|p_{jm}\right|^{2}-\Re\left(a_{j}p_{jm}p_{jm}^{*}\right)\right]\nonumber \\
 & = & -\frac{I_{0}}{c}\frac{\lambda^{3}}{8\pi^{3}}\left[\left|a_{j}\right|^{2}-\Re\left(a_{j}\right)\right]\sum_{m=-j}^{j}m\left|p_{jm}\right|^{2}.\label{eq:Torque}
\end{eqnarray}

Considering only an electric dipole MC, the relation simplifies to:

\begin{eqnarray}
N_{\mathrm{p}} & = & -\frac{I_{0}}{c}\frac{\lambda^{3}}{8\pi^{3}}\left[\left|a_{1}\right|^{2}-\Re\left(a_{1}\right)\right]\left(\left|p_{11}\right|^{2}-\left|p_{1-1}\right|^{2}\right),
\end{eqnarray}

\noindent which suggests that a circularly polarized plane wave (Eq.~\ref{eq:Cpo}),
where all the power is put into one amplitude, is the optimum illumination
to exert maximal torque on an absorbing isotropic particle.

Considering higher order MCs and assuming a circularly polarized plane
wave illumination, using equation Eq.~\ref{eq:Cabs}, Eq.~\ref{eq:Torque}
simplifies to:

\begin{eqnarray}
N_{j^{\mathrm{e}}} & = & \sigma4\left(2j+1\right)\left[\Re\left(a_{j}\right)-\left|a_{j}\right|^{2}\right]N^{\mathrm{norm}}=\frac{\sigma I_{0}C_{\mathrm{abs},j^{\mathrm{e}}}}{\omega},
\end{eqnarray}

\noindent and the maximum torque, occurs for the maximum absorption
cross section (Eq.~\ref{eq:CabsMax}) and reads as:

\begin{eqnarray}
\left[N_{j^{\mathrm{e}}}\right]_{\mathrm{max}} & = & \frac{\sigma I_{0}\left[C_{\mathrm{abs},j^{\mathrm{e}}}\right]_{\mathrm{max}}}{\omega}=\sigma\left(2j+1\right)N^{\mathrm{norm}}.
\end{eqnarray}

Similarly, for a magnetic MC $b_{j}$, the following relation derives:

\begin{eqnarray}
\left[N_{j^{\mathrm{m}}}\right]_{\mathrm{max}} & = & \sigma\left(2j+1\right)N^{\mathrm{norm}}.
\end{eqnarray}

\subsection*{Fundamental bounds for the total optical torque}

Calculating the maximum of the total optical torque is easier than the force, since no interference term contributes to the torque along the wave propagation. To maximize the torque, it is required to overlap all the MC resonances spectrally and 
tune the loss for critical coupling at that frequency.  Therefore, the maximum total optical torque as a
function of $j_\mathrm{max}$ is only a summation of all individual contributions $\sum_{j=1}^{j_\mathrm{max}}2(2j+1)N^{\mathrm{norm}}=2j_\mathrm{max}(j_\mathrm{max}+2)$. This relation is plotted in Fig.~3(b).

\subsection*{Cartesian dipole expansion}

The time averaged optical torque exerted on a dipolar particle by
an arbitrary incident wave, expanded into Cartesian dipole moments,
is (SI units) \cite{National2015a}:

\begin{eqnarray}
\mathbf{N} & = & \frac{1}{2}\biggl\{\Re\left(\mathbf{p}\times\mathbf{E}_{\mathrm{inc}}^{\mathrm{*}}+\mathbf{m}\times\mathbf{B}_{\mathrm{inc}}^{\mathrm{*}}\right)\label{eq:Torque_moments}\\
 &  & -\frac{k^{3}}{6\pi}\left[\frac{1}{\epsilon_{0}}\Im\left(\mathbf{p}^{*}\times\mathbf{p}\right)+\mu_{0}\Im\left(\mathbf{m^{*}}\times\mathbf{m}\right)\right]\biggr\}.\nonumber 
\end{eqnarray}

\section{Particle Swarm Optimization (PSO)}

Particle swarm optimization (PSO) is a search-based optimization method, based on motion and intelligence
of swarms, in which individuals orient their motion towards the
personal ($P_\mathrm{best}$) and overall ($G_\mathrm{best}$) best locations imposed
by an associated objective function of particles.
The modification of particles is defined through the velocity concept
as \cite{lee2008modern}:

\noindent 
\begin{eqnarray}
\nu_{i}^{it+1} & = & w^{it}\nu_{i}^{it}+c_{1}\times rand\times\left(P_{\mathrm{best}\, i}-s_{i}^{it}\right)+c_{2}\times rand\times\left(G_{\mathrm{best}\, i}-s_{i}^{it}\right),\\
s_{i}^{it+1} & = & s_{i}^{it}+\nu_{i}^{it+1},\\
w^{it} & = & w_\mathrm{max}-\left(\frac{w_\mathrm{max}-w_\mathrm{min}}{iter_\mathrm{max}}\right)\times it,
\end{eqnarray}
\noindent where $\nu_{i}^{it}$ and $s_{i}^{it}$ are the velocity
and position of particle $i$ at PSO iteration $it$, $rand$ is a
random number between 0 and 1, and $w$ is the weighting factor. The
values $c_{1}=c_{2}=2$, $w_\mathrm{max}=0.9$, and $w_\mathrm{min}=0.4$ are suggested
as proper input values, independent of the problem \cite{lee2008modern}. We have used PSO To maximize the optical force exerted on a core-multishell particle in dipolar approximation. The variables are the three radii of the core and shells and the 
objective function is ${F}/{F^\mathrm{norm}}$.
The result is shown in Fig.~3(c).

\section{Numerical Results}

\begin{center}
\begin{figure}
\begin{centering}
\includegraphics[scale=0.4]{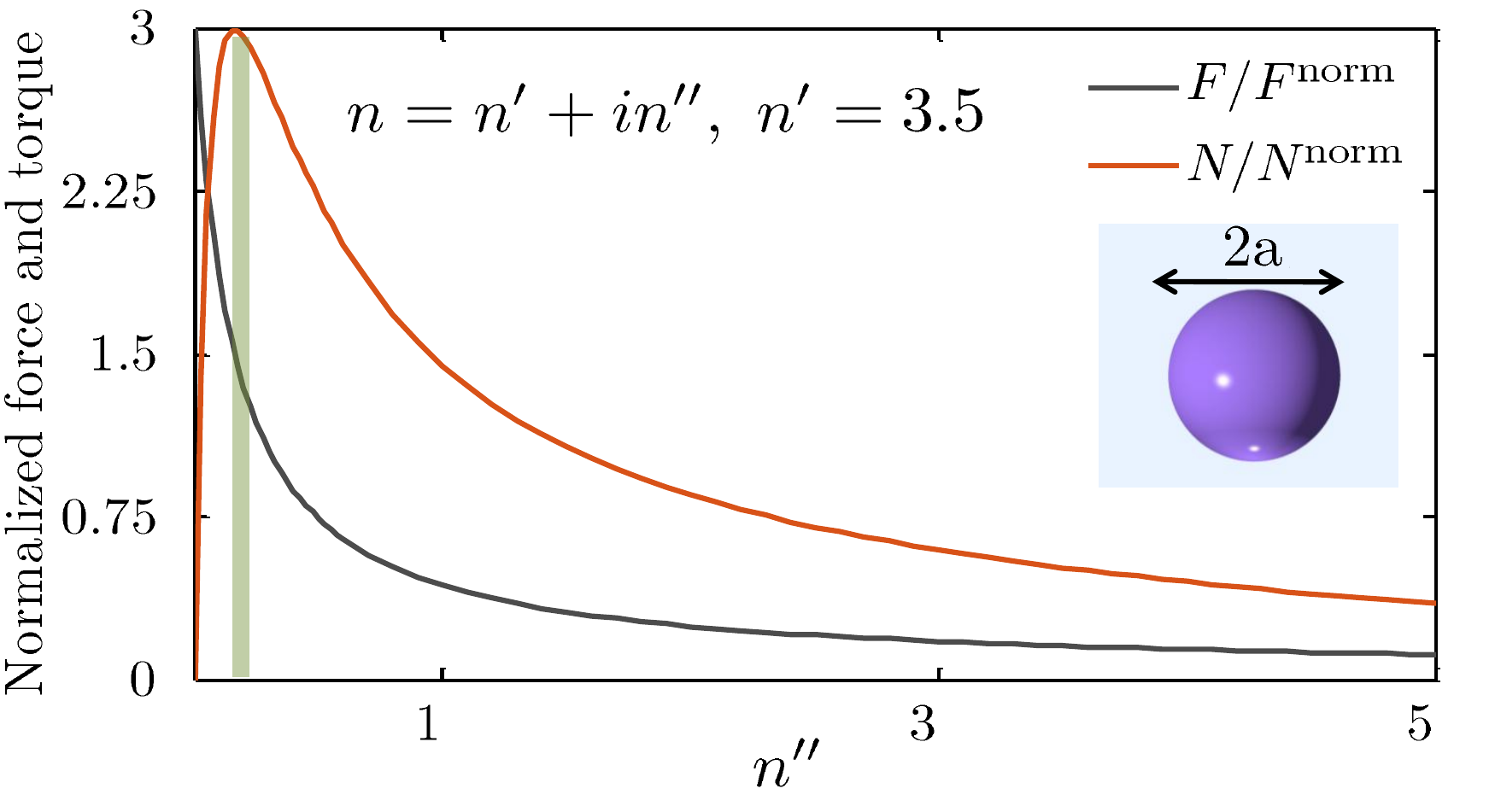}
\par\end{centering}
\protect\caption{a) The normalized maximal force and torque exerted by an arbitrarily
polarized plane wave (for the force) or a circularly polarized plane
wave (only for the torque) at the resonance of a dipole MC on an isotropic
particle, as a function of the imaginary part of the refractive index.
Refractive index is $n=n'+\mathrm{i}n^{\prime\prime}$. Maximum torque occurs at
the critical coupling regime (shown by a green strip), i.e $\gamma_{nr}\approx\gamma_{r}$. (A
linearly polarized plane wave exerts no torque on the particle.) \label{Imaginary}}
\end{figure}
\par\end{center}

\begin{center}
\begin{figure}
\begin{centering}
\includegraphics[scale=0.1]{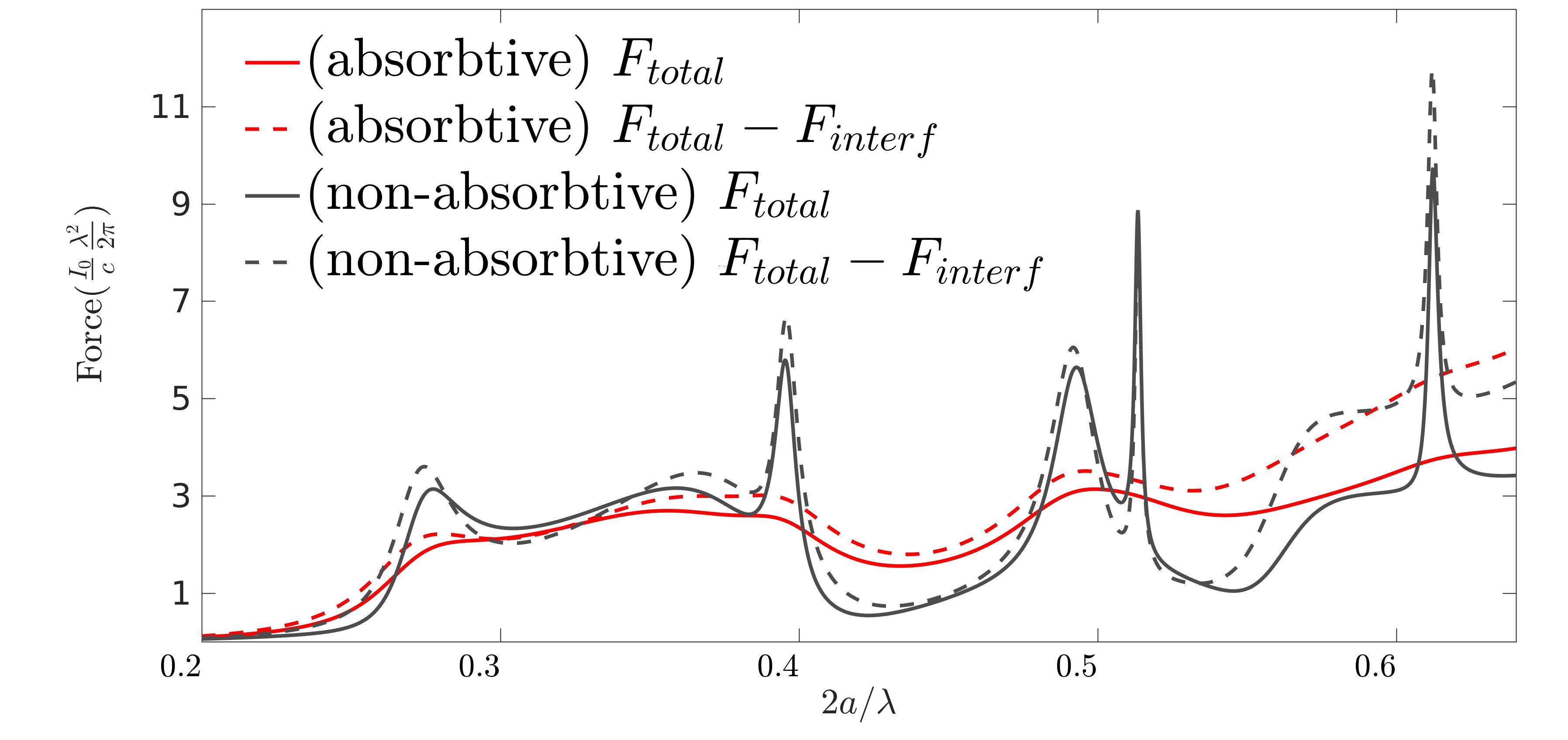}
\par\end{centering}
\protect\caption{Time averaged total optical force exerted by an arbitrarily polarized plane wave on a non-absorbing sphere ($\epsilon_\mathrm{r} = (3.5)^2$, $\mu_\mathrm{r}=1$) (black) and an absorbing sphere ($\epsilon_\mathrm{r} = (3.5+0.16\mathrm{i})^2$, $\mu_\mathrm{r}=1$) (red) as a function of
the sphere's size parameter (solid line). The contribution of the non-interference ($=F_{total}-F_{interf}$) terms (dashed line). \label{dissipation}}
\end{figure}
\par\end{center}

\begin{figure}
\begin{centering}
\includegraphics[scale=0.4]{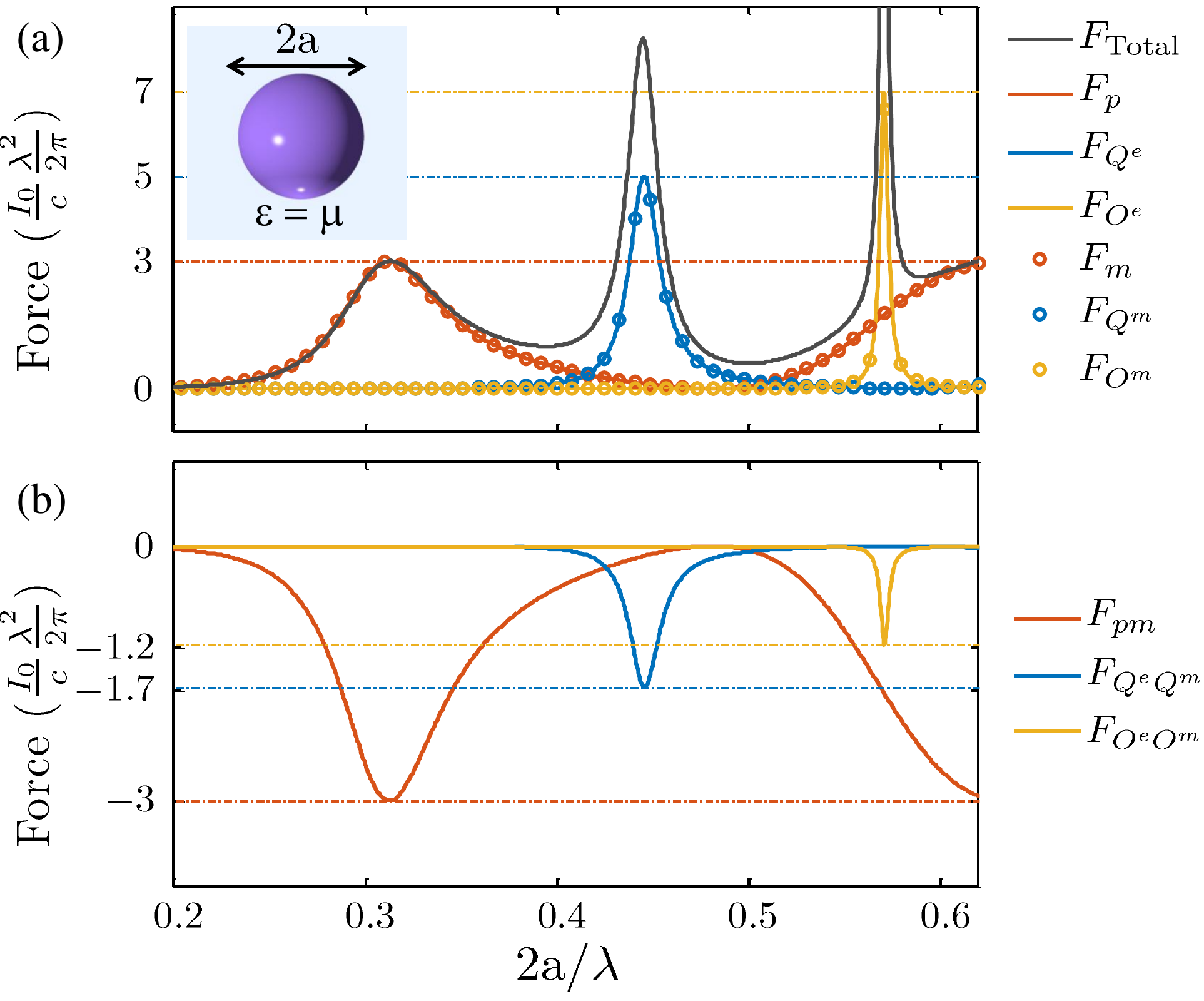}
\par\end{centering}
\protect\caption{\textit{Non-absorbing dual dielectric sphere}: a) The time averaged total optical force exerted on a non-absorbing dual sphere ($\epsilon_\mathrm{r} = \mu_\mathrm{r}= 3.5$),
illuminated by an arbitrarily polarized plane wave, as a
function of the sphere's size parameter. The individual
contribution of the dipole, quadrupole, and octopole electric
and magnetic MCs. b) The partial contribution due to the interference of dipole-dipole,
quadrupole-quadrupole, and octopole-octopole electric and magnetic
MCs. \label{fig:Non-absorptive_dual_dielectric}}
\end{figure}

\begin{center}
\begin{figure}
\begin{centering}
\includegraphics[scale=0.4]{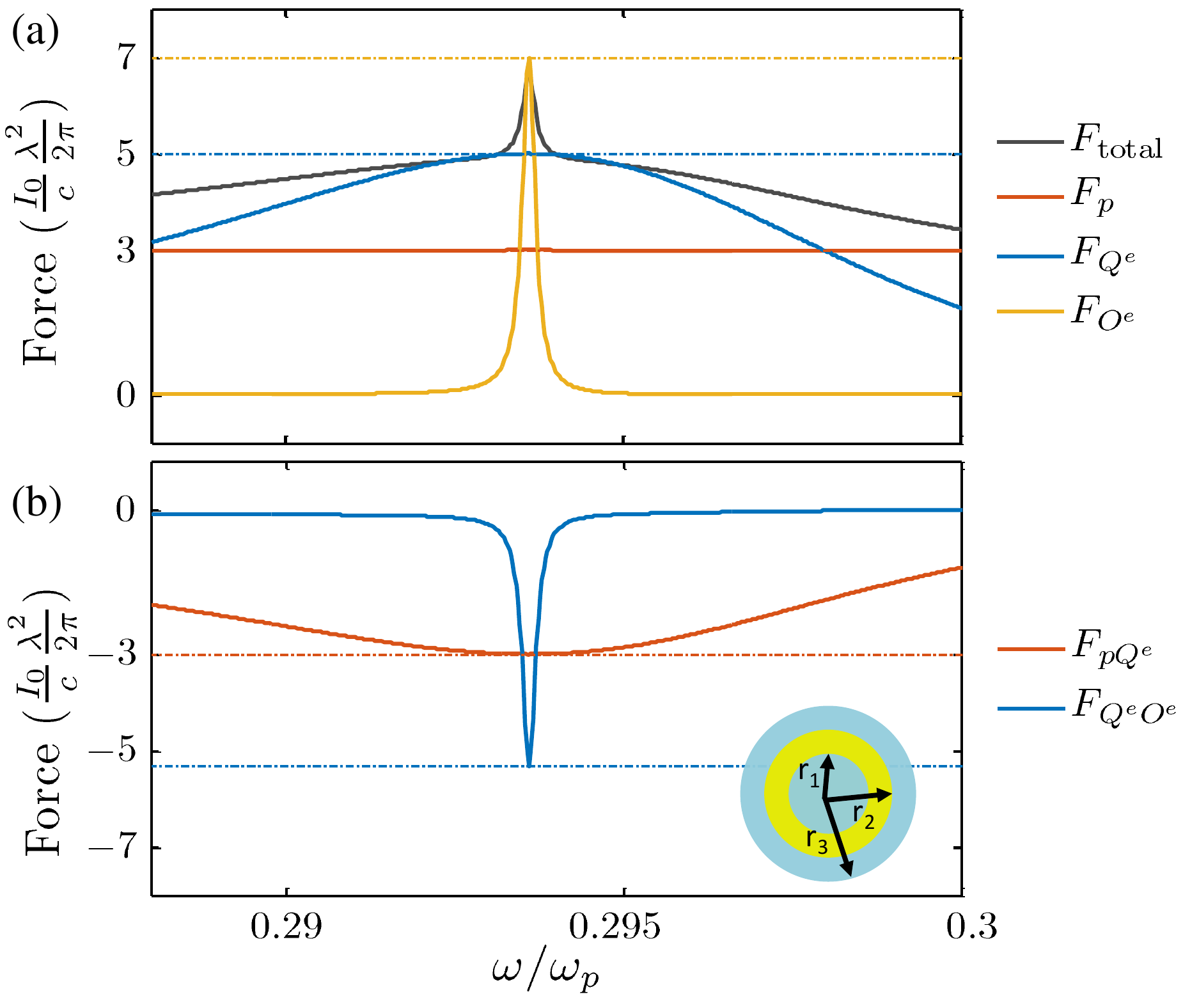}
\par\end{centering}
\protect\caption{\textit{Non-absorbing core-multishell particle}: a) The time averaged total optical
force exerted on an isotropic non-absorbing silver-silicon-silver core-multishell particle, illuminated
by an arbitrarily polarized plane wave, as a function of the normalized
frequency $\omega/\omega_{\mathrm{p}}$ and the non-negligible partial
contributions of the individual dipolar, quadrupolar, and octopolar
electric MCs; and b) the interference of dipole-quadrupole and quadrupole-octopole,
electric MCs. The upper bounds for the contributions are shown by
a dashed line of the same color. ($r_{1}=0.4749\lambda_{\mathrm{p}},\, r_{2}=0.6404\lambda_{\mathrm{p}},\, r_{3}=0.8249\lambda_{\mathrm{p}}$,
$\lambda_{\mathrm{p}}=2\pi c/\omega_{\mathrm{p}},\,\epsilon_{\mathrm{silicon}}=12.96$
\cite{Ruan2011a}) \label{fig: multi-shell}}
\end{figure}
\par\end{center}

\begin{center}
\begin{figure}
\begin{centering}
\includegraphics[scale=0.4]{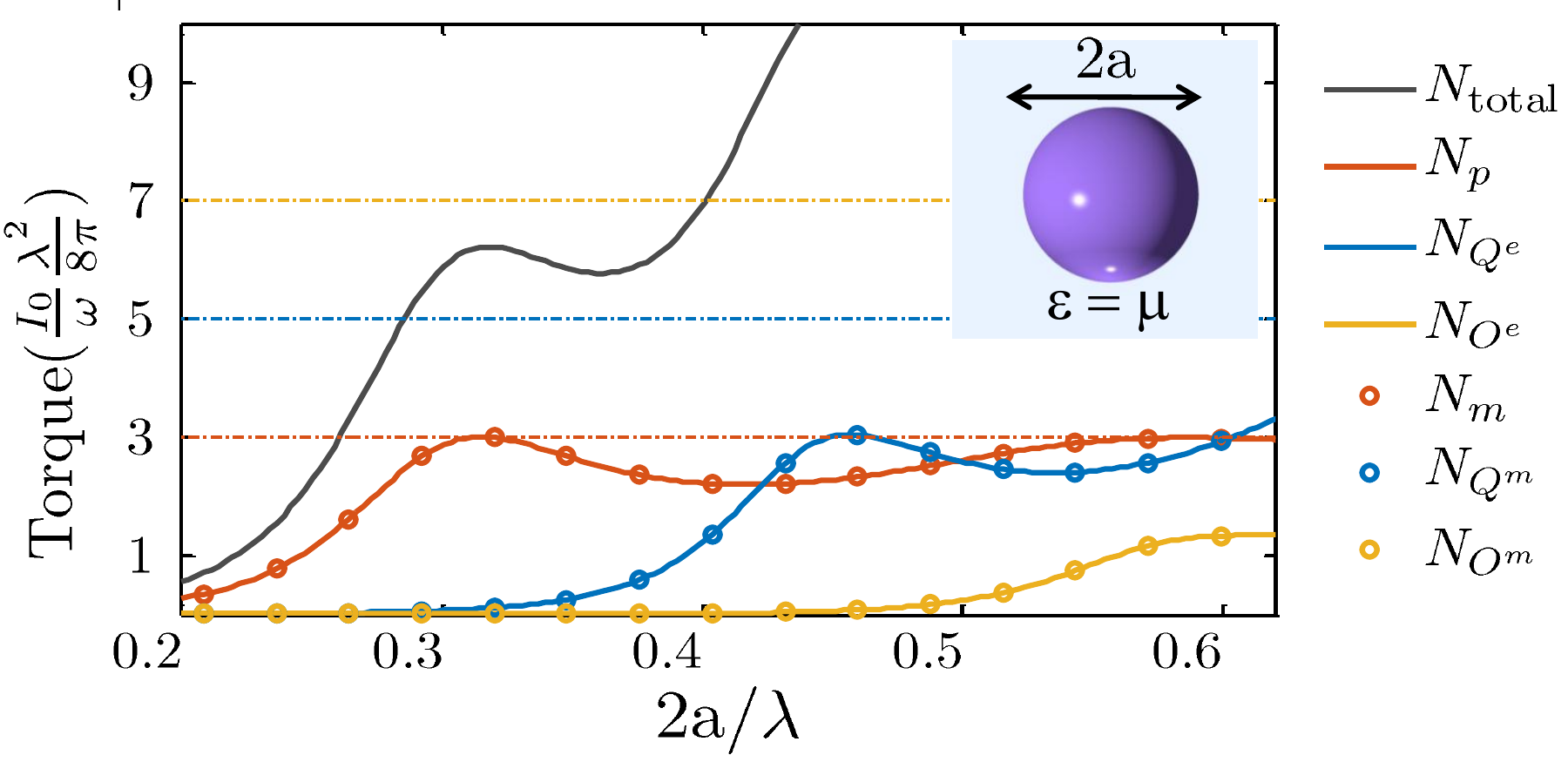}
\par\end{centering}

\protect\caption{\textit{Absorbing dual dielectric sphere}: The time averaged total
optical torque exerted on an absorbing dual sphere,
illuminated by a circularly polarized plane wave as a function of
the sphere's size parameter. The individual contribution of the
dipole, quadrupole, and octopole electric and magnetic MCs. The maximum contribution
 for each term is shown by a dashed line of the same
color. \label{fig:Absorptive_dual_dielectric}}
\end{figure}
\par\end{center}

\begin{center}
\begin{figure}
\begin{centering}
\includegraphics[scale=0.4]{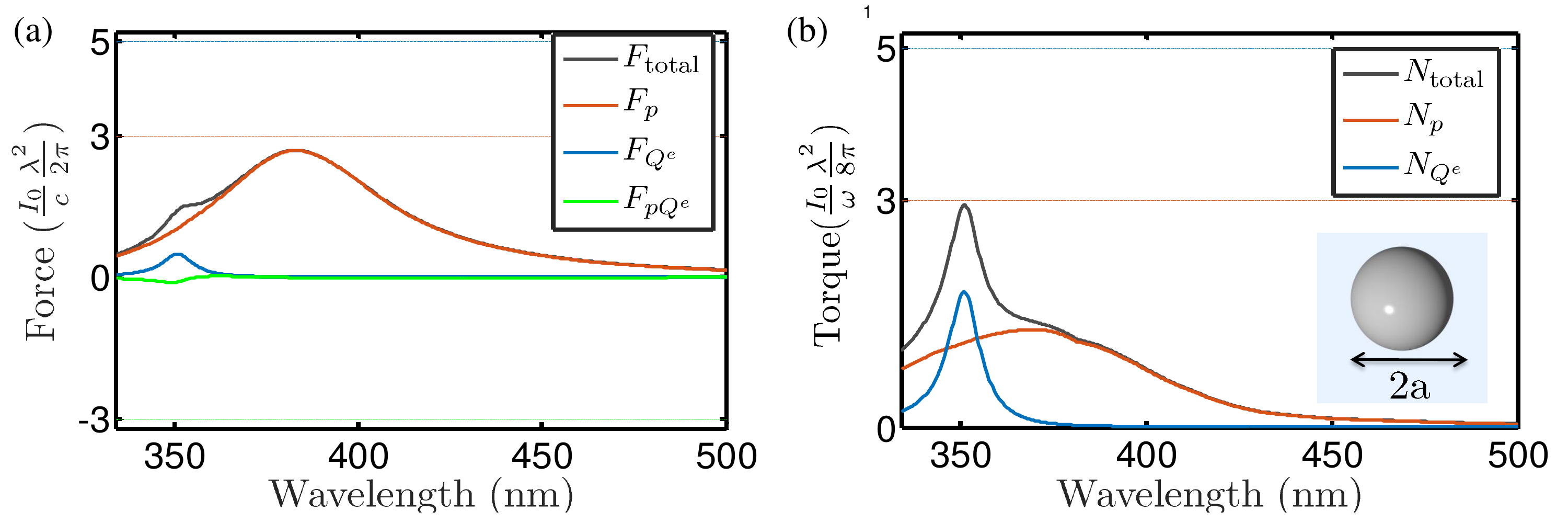}
\par\end{centering}

\protect\caption{\textit{Silver sphere}: a) The time averaged total
optical force and b) torque exerted on a silver sphere ($a=45$ nm), illuminated by an arbitrary polarized plane wave (for the force) and a circularly polarized plane wave (only for the torque). The individual contribution of the
dipole and  quadrupole electric MCs and their interference. The maximum contribution
 for each term is shown by a line of the same
color. (Experimental data from Ref.~\onlinecite{hagemann1975optical}).
 \label{fig:Silver_Sphere_Force_Torque}}
\end{figure}
\par\end{center}


Figure~\ref{Imaginary}, explores the effect of the imaginary part
of the particle's material refractive index (related to non-radiative
loss) on the maximal force and torque (i.e., at the MC resonance)
exerted by an arbitrarily polarized plane wave on a single channel
dipolar particle.

In Fig.~2a, we have plotted the total force and the non-interference contributions to the force for a non-absorbing dielectric sphere. To see the effect of absorption, in Fig.~\ref{dissipation}, we have also plotted the total force and the non-interference terms for an absorbing sphere. The figure helps to compare the effect of absorption on the interference terms and the total force.

In Fig.~\ref{fig: multi-shell}, the optical force exerted by an
arbitrarily polarized plane wave on a silver-silicon-silver core-multishell particle
is calculated. The metal is characterized by a lossless Drude model
$\epsilon_{\mathrm{metal}}=1-\omega_{p}^{2}/\omega^{2}$, where $\omega_{p}$
is the plasma frequency of the silver. The electric dipole-quadrupole
and quadrupole-octopole MC interference force contributions are engineered
to be maximized by spectrally overlapping the electric dipole ($a_{1}$),
quadrupole ($a_{2}$) and octopole ($a_{3}$) MC resonances. The particle
is primarily designed for the purpose of supper-scattering \cite{Ruan2011a}.
Unlike optical cross sections, the total optical force is not derived
by adding the individual MC contributions. However, the interference
terms contribute to the total optical force significantly. Such that
the total force is equal to the individual contribution of a single
electric octopole.

In Fig.~\ref{fig:Absorptive_dual_dielectric}, the optical torque exerted by a
circularly polarized plane wave on a dual dielectric sphere is calculated. Due to duality, the electric and magnetic MCs are identical. In the analytical relation for optical torque, no interference term appears. Therefore, the maximum total exerted torque on the dual sphere for one multipole is twice as big as $(N_{j^\mathrm{e}})_{\mathrm{max}}$. This makes dual particles as good candidates for super-rotatable objects.

In Fig.~\ref{fig:Silver_Sphere_Force_Torque}, the optical force and torque on a realistic silver nanoparticle is calculated to be able to compare the results with the previous results. The interesting observation is that the electric quadrupolar contribution is much more pronounced in the exerted optical torque than in the force. This large torque is due to the fact that the particle absorbs considerably at the electric quadrupole resonance.

